\documentclass[twocolumn,aps,showpacs,prb,tightenlines,amsmath,amssymb,superscriptaddress]{revtex4} 
                       
\usepackage{graphicx}    
\usepackage{bm}
\usepackage{colordvi}                        
\usepackage{color}
\usepackage[normalem]{ulem}

\def\up{\uparrow}
\def\down{\downarrow}

\def\ran{\rangle}
\def\lan{\langle}
\def\fr{\frac}
      
\begin{document}             
\title{Spin-dependent inter- and intra-valley electron-phonon
 scattering in germanium}                    
\author{Z. Liu}
\affiliation{Hefei National Laboratory for Physical Sciences at Microscale and
  Department of Physics, University of Science and Technology of China, Hefei,
  Anhui, 230026, China} 
\author{M. O. Nestoklon}     
\affiliation{A. F. Ioffe Physico-Technical Institute, Russian Academy of Sciences, St. Petersburg 194021, Russia} 
\author{J. L. Cheng}     
\affiliation{Department of Physics and Institute for Optical Sciences,
University of Toronto, 60 St. George Street. Toronto, Ontario, Canada
M5S 1A7} 
\author{E. L. Ivchenko}  
\email{ivchenko@coherent.ioffe.ru}  
\affiliation{A. F. Ioffe Physico-Technical Institute, Russian Academy of Sciences, St. Petersburg 194021, Russia} 
\author{M. W. Wu}   
\email{mwwu@.ustc.edu.cn.}
\affiliation{Hefei National Laboratory for Physical Sciences at Microscale and
  Department of Physics, University of Science and Technology of China, Hefei,
  Anhui, 230026, China}
\date{\today}

\begin{abstract}
  We investigate the
  spin-dependent electron-phonon scatterings of the $L$ and $\Gamma$
  valleys and the band structure near the
  conduction band minima in
germanium. We first construct a $16\times16$ ${\bm k}\cdot{\bm p}$
  Hamiltonian in the vicinity
  of the $L$ point in germanium, which ensures the correctness of the band
structure of the lowest three conduction bands and highest two valence
  bands. This Hamiltonian facilitates the 
  analysis of the spin-related properties of the conduction electrons.  
We then demonstrate the
  phonon-induced electron scatterings of the $L$ and $\Gamma$
  valleys, i.e., the  intra-$\Gamma$/$L$ valley, 
inter--$\Gamma$-$L$ valley and inter--$L$-$L$
  valley scatterings in germanium. The selection rules and complete scattering
  matrices for these scatterings are calculated, where
  the scattering matrices for the
  intra-$\Gamma$ valley scattering, inter--$\Gamma$-$L$ valley scattering and the
  optical-phonon and the separated transverse-acoustic-
 and longitudinal-acoustic-phonon contributions to the intra-$\Gamma$
 valley scattering have not been reported in the literature.
The coefficients in these scattering matrices are
 obtained via the pseudo-potential calculation, which also verifies our
 selection rules and wave-vector dependence. We further discuss the
 Elliott-Yafet mechanisms in these electron-phonon scatterings with
 the ${\bm k}$$\cdot$${\bm p}$ eigenstates at the $L$ and $\Gamma$
 valleys. Our investigation of these electron-phonon scatterings are
 essential for the study of the optical orientation of spin and 
 hot-electron relaxation in germanium.
\end{abstract}


\maketitle
\section{INTRODUCTION}

The group IV materials are attractive and promising candidates for the
achievement of spintronic devices.\cite{Zutic-Si1,PeterYu1,Gray-Si1,Jeon-Si1,Jonker-Si1,Appelbaum-Si1,Grenet-Si1,Dash-Si1,Culcer-Si1,Loren-Ge1,Loren-Ge2,Pezzoli-Ge1,Rioux-Ge1,Li-Ge1,Tang-Ge1,Li-Ge2,Jain-Ge2,Jain-Ge1,Li-Ge3,Zhou-Ge1} The Dyakonov-Perel mechanism is
absent in these materials due to the
 centrosymmetry and the hyperfine interaction can be
suppressed by isotopic purification, which ensures a relative long
spin-decoherence
time.\cite{Culcer-Si1,Loren-Ge1,Loren-Ge2,Zutic-Si1,Tahan1,Sherwin1,Shor1,Bennett1,Cory1}
Also, the silicon-based microfabrication
technology  is well-developed and extensively used.\cite{Culcer-Si1} 
Germanium (Ge), as a group IV element adjacent to silicon, 
shares the good spin-decoherent property
and is fully compatible with the existing mature nanoelectronic
technology in silicon (Si).\cite{PeterYu1,Loren-Ge1,Loren-Ge2,Pezzoli-Ge1}
Particularly, in contrast to Si, Ge shows obvious
electro-optic effect as  its direct gap (at the $\Gamma$ point)
is close to the indirect
gap (at the $L$ point) and lies in the infrared
range.\cite{Loren-Ge2,Pezzoli-Ge1,Rioux-Ge1,Li-Ge1,Tang-Ge1,Li-Ge2,Li-Ge3,Jain-Ge2,Jain-Ge1,Cameron-optical1,Jancu1} Thus the optical orientation 
of carriers, which is free from the interfacial effect and
the external electric and magnetic fields, can be
carried out effectively in Ge-based
devices.\cite{Loren-Ge2,Pezzoli-Ge1,Rioux-Ge1,Kuo-Nature-437-1334} Moreover, compared with Si,  the longer spin-diffusion length stemming from the
larger carrier mobility is helpful to the spin injection and the
relative  strong spin-orbit coupling benefits
the manipulation of spin.\cite{Loren-Ge1,Loren-Ge2,Jain-Ge2,Jain-Ge1,Pezzoli-Ge1}

In recent years, Ge attracts a renewed interest both
experimentally and theoretically. In the experimental side,
  Loren {\em et al.}\cite{Loren-Ge1,Loren-Ge2} and Pezzoli {\em et al.}\cite{Pezzoli-Ge1} demonstrated the optical injection and detection of polarized electrons and holes in bulk Ge and Ge-based quantum
wells, where
electrons are pumped optically in the $\Gamma$ valley and quickly scattered
to the indirect valleys.  In the theoretical side, a progress was made 
in investigations of the conduction band structure and spin-dependent 
electron-phonon
scattering.\cite{Rioux-Ge1,Li-Ge1,Tang-Ge1,Li-Ge2} A compact 
10$\times$10 ${\bm k}\cdot{\bm p}$ Hamiltonian was constructed
 around the $L$ point via the method of
invariant.\cite{Li-Ge2,Group1,Voon1,Bir1}  Moreover, Tang {\em  et al.}\cite{Tang-Ge1} derived the selection rules for intra- and 
inter-$L$ valley electron-phonon scattering, and calculated the average
absolute values of corresponding scattering elements within
 the tight-binding model.  Later Li {\em et al.}\cite{Li-Ge2} 
demonstrated the scattering
matrices for the inter-$L$ valley scattering and acoustic
(AC) contribution to the intra-$L$ valley scattering by using the
  pseudo-potential method, where the approximated wave-vector
dependence of the intra-$L$ valley scattering is derived with the
combination of ${\bm k}$$\cdot$${\bm p}$, pseudo-potential and
group theories.\cite{Dery_Si1} Very recently Li {\em et al.}\cite{Li-Ge3} also
  reported the selection rules of the intra--$\Gamma$-$L$ valley electron-phonon
  scattering in the calculation of phonon-assisted optical transitions
  in Ge. However, to our knowledge, the complete scattering
  matrices of several important channels of the electron-phonon
scattering have not been discussed yet, such as  the intra-$\Gamma$ valley
  scattering, the optical-phonon (OP)
 contribution to the intra-$L$ scattering as
   well as the inter--$\Gamma$-$
 L$ valley scattering. As shown in previous works, these scatterings are
 fundamental to understand the spin dynamics in the optical
 orientation of electron spin in
  Ge.\cite{Rioux-Ge1,Loren-Ge1,Loren-Ge2,Pezzoli-Ge1,Li-Ge3}  
 
In this work, we readdress the band structure of Ge
near the conduction  band  minima and study the phonon-induced electron scatterings in the $L$ and $\Gamma$ valleys.
We first derive a spin-dependent 16$\times$16 ${\bm k}$$\cdot$${\bm p}$
Hamiltonian in the vicinity of the $L$ point in Ge from the 
band basis functions at this
 point.\cite{PeterYu1,Voon1} Compared with Ref.~\onlinecite{Li-Ge2}
 where the effective electron masses are taken as granted, 
here we start from the free electron
mass and straightforwardly obtain the renormalization of masses 
from this Hamiltonian.  It can fit the band structure of
 the lowest three conduction bands and highest two valence bands, and provides the eigenstates for quantitative
 demonstration of conduction electron spin properties.

Till now,  in  addition to the well-known subgroup technique which gives the
   selection rules of the wave-vector-independent contribution to the electron-phonon interaction,\cite{Lax-Selection1,Lax2,Yafet1,Tang-Ge1} two  approaches for
 deriving the explicit wave-vector dependence of
   scattering matrix have been brought
   forward.\cite{Dery_Si1,Li-Ge2,McCann-phonon1,Ochoa-phonon1}
In one approach the initial and final electronic states are expressed as the
 ${\bm k}\cdot{\bm p}$ eigenstates. The wave-vector dependence of
 scattering matrix element is the product of that in the electron/phonon
 states and crystal potential.\cite{Dery_Si1,Li-Ge2} Here
 the selection rule for each analytical term is given by the group-theory
 analysis from the symmetries of the ${\bm k}$$\cdot$${\bm p}$ basis
 functions, phonon state and  crystal potential, while  the
 coefficients of the electron-phonon scattering matrices are integrals involving the ${\bm k}$$\cdot$${\bm p}$ basis
 functions and crystal potential  and can be
 calculated straightforwardly via the pseudo-potential method. The other approach, {i.e.}, the theory of invariants,  utilizes
 the invariance of electron-phonon interaction to the symmetry
 operators in the corresponding space group.\cite{Voon1,Bir1}
The invariant scattering matrix consists of products of
wave-vector-dependent irreducible tensor 
components and the bare spin-dependent matrices
and takes into account the symmetry of phonon states. The corresponding
 coefficients should be
obtained via calculation with numerical techniques such as 
 pseudo-potential or tight-binding methods.\cite{Group1,Voon1,Bir1,McCann-phonon1,Ochoa-phonon1}
 Obviously, the validity of the first approach is sensitive to the
 choice of ${\bm k}$$\cdot$${\bm p}$ eigenstates, and the
 second one is not limited by the ${\bm k}$$\cdot$${\bm p}$
 Hamiltonian. Therefore we take the invariant method, and determine the
 coefficients via fitting with our pseudo-potential calculations.

By applying the method of invariants, we construct the scattering matrices and investigate the intra- and inter-valley scatterings involving both
the $\Gamma$ and four $L$ valleys. The matrix elements for the intra-$\Gamma$ and inter--$\Gamma$-$L$ valley scattering,
 the OP contribution and separated transverse acoustic (TA)
 and longitudinal acoustic (LA) contributions to the intra-$L$
 valley electron-phonon scattering are provided for the first
 time. For each phonon mode,
   we demonstrate the lowest-order wave-vector dependence of the
   scattering matrix as it is much larger than the higher-order
   terms. It should be noted that the zeroth-order contribution to the spin-flip
   scattering elements does exist in the inter-valley 
scattering but vanishes in the
   intra-valley case. 
 Furthermore, with our ${\bm k}$$\cdot$${\bm p}$ Hamiltonian at the $L$ point and a $14\times14$ ${\bm k}$$\cdot$${\bm p}$ Hamiltonian at the $\Gamma$ point, we analyze the Elliott\cite{Elliott1} and Yafet\cite{Yafet1} mechanisms 
in these electron-phonon scatterings.\cite{Ridene-Gamma1}
 In all the cases above, our  calculations with pseudo-potential
 method confirm our selection rules and
  analytical wave-vector dependence of scattering matrices.

This paper is organized as follows. In Sec.\,II we construct the
$16\times16$ ${\bm k}$$\cdot$${\bm p}$ Hamiltonian. In Sec.\,III, we investigate
the mechanisms of the phonon-induced electron scattering, where
the intra-$\Gamma$/$L$ valley scattering, inter--$\Gamma$-$L$ valley scattering
 and inter--$L$-$L$ valley  scattering are studied analytically via the symmetry
 consideration. Besides, we calculate the
 scattering matrices numerically
 with pseudo-potential method.  We summarize in Sec.\,IV.\@

\section{THE ${\bm k}$$\cdot$${\bm p}$ HAMILTONIAN}
Around the conduction band minimum of Ge, the ${\bm k}$$\cdot$${\bm p}$
Hamiltonian with the spin-orbit coupling included 
can be derived from the symmetry
at the four $L$ points $(\pi/a)(1,1,1)$, $(\pi/a)(-1,-1,1)$,
$(\pi/a)(-1,1,1)$ and $(\pi/a)(1,-1,1)$ with $a$ being the lattice constant. 
At each point, the symmetry of the Bloch states
 is described by the $D_{3d}$ double
group with the six irreducible representations $L^{+}_4$, $L^{+}_5$,
 $L^{+}_6$,  $L^{-}_4$,  $L^{-}_5$ and
$L^{-}_6$.\cite{Group1,PeterYu1,Winkler1} We first
consider the vicinity of the $(\pi/a)(1,1,1)$ point. For the sake 
of convenience, we choose the coordinate
system $x,y,z$ with the $z$ direction along the symmetry axis
 [111], so that the unit vectors of this system,  related
with those of the crystallographic frame $[100]$, $[010]$ and $[001]$,
i.e.,  $\Hat{x}_0$, $\hat{y}_0$ and $\hat{z}_0$,  
by $\hat{x}=(\hat{x}_0-\hat{y}_0)/\sqrt{2}$, $\hat{y}=(\hat{x}_0+\hat{y}_0-2\hat{z}_0)/\sqrt{6}$ and  $\hat{z}=(\hat{x}_0+\hat{y}_0+\hat{z}_0)/\sqrt{3}$. 
Starting from the nonrelativistic Bloch states
 $L_1, L_3, L^c_{3'}, L_{2'}$ in the conduction band and the 
$L_{3'}^v$ states in the valence band and including the
 spin-orbit interaction, we obtain 16 Bloch states 
given in Table~\ref{tableX} of Appendix~\ref{appA}, which are
  used in the construction of the  ${\bm k}$$\cdot$${\bm p}$ Hamiltonian
  and the analysis of the electron-phonon scattering.

Due to the space inversion symmetry in bulk Ge,
the basis functions have defined parities. For this reason, 
 the superscripts `$+$' and
`$-$' are used to represent the even and odd parities, respectively. 
It is noted that all
the bands are 2-fold degenerate due to the time inversion 
symmetry.\cite{PeterYu1} The spin-dependent perturbation Hamiltonian is 
given by\cite{Winkler1,dymnikov}
\begin{equation}
\Delta H = \fr{\hbar}{4m^2_{\rm e}c^2} \left[ \bm \nabla
V({\bm r})\times{\bm p}\right]\cdot{\bm \sigma} + \fr{\hbar{\bm
    k}\mbox{$\cdot$}\bm \pi}{m_{\rm e}}+ \fr{\hbar^2k^2}{2m_{\rm e}}\:,
\end{equation}
in which ${\bm k}$ is the electron wave vector referred to the $L$-point, the 
first term describes the spin-orbit interaction at ${\bm k} = 0$, 
$m_e$ is the free electron mass, $V({\bf r})$ is the spin-independent
periodic potential and ${\bm \pi}$ is the generalized momentum operator \cite{dymnikov} 
\begin{equation} \label{pip}
{\bm \pi} = {\bm p} + \delta {\bm \pi}\:,\: \delta {\bm \pi} = \frac{\hbar}{4 m_e c^2} {\bm \sigma} \times {\bm \nabla} V({\bm r})\:. 
\end{equation}
The total ${\bm k}$$\cdot$${\bm p}$ Hamiltonian matrix is written 
in the form of three terms
\begin{equation}
H = \frac{\hbar^2k^2}{2m_{\rm e}} \hat{I} + H_{0} + H_{{\bm k}{\bm p}} \:, 
\label{Htotal2}
\end{equation}
where $H_{0}$ is the Hamiltonian matrix at the $L$-point
and the $H_{{\bm k}{\bm p}}$ is the linear-${\bm k}$ contribution 
describing the interband ${\bm k}$$\cdot$${\bm \pi}$ mixing.
The diagonal components of $H_{0}$ are introduced in the 
third column of Table~\ref{tableX} of Appendix~\ref{appA}.
The matrix $H_0$ also has off-diagonal components responsible for the
 interband spin-orbit mixing which takes place only between
the band states transforming according to the
equivalent spinor representations. The nonzero off-diagonal components
which stem from $(\nabla V\times {\bf p})\cdot{\bm \sigma}$ terms  are
\begin{subequations}
\begin{align}
\langle v4 \vert H_0 \vert c12 \rangle &= \langle v3 \vert H_0 \vert c11 \rangle = \Delta_1\:,\\
\langle c1 \vert H_0 \vert c4 \rangle &= \langle c2 \vert H_0 \vert c3 \rangle = \Delta_2\:, \\
\langle c9 \vert H_0 \vert c11 \rangle &= \langle c10 \vert H_0 \vert c12 \rangle = \Delta_3\:, \\
 \langle v4 \vert H_0 \vert c10 \rangle &= \langle v3 \vert H_0 \vert c9 \rangle 
= - \langle v2 \vert H_0 \vert c8 \rangle \nonumber \\&= - \langle v1 \vert H_0 \vert c7 \rangle = \Delta_4 
\end{align}
\end{subequations}
together with 10 transposed matrix elements. Here $\Delta_l$ $(l$ = 1, 2, 3, 4)
are real band parameters and are listed in Table~\ref{table1}.

The linear-${\bm k}$ matrix can be rewritten as
\begin{eqnarray}
H_{{\bm k}{\bm p}} =  \left[ \begin{array}{cc}
    H_{\rm cc}&H_{\rm vc}^{\dagger}\\
    H_{\rm vc}&0
  \end{array}\right]\:,
\label{Htotal}
\end{eqnarray}
where $H_{\rm cc}$ and $H_{\rm vc}$ are 12$\times$12 and 4$\times$12 block submatrices, respectively. Taking into account that the matrix elements between the states of coinciding parities vanish, these submatrices can further be presented in the form
\begin{eqnarray}
H_{\rm cc} =  \left[ \begin{array}{cc}
    0&H_{\rm cc}^{+-}\\
    H_{\rm cc}^{-+}&0
  \end{array}\right]\:,\:~~ H_{\rm vc} =  \left[ \begin{array}{cc}
       0&H_{\rm vc}^{-+}   \end{array}\right]\:,
\label{Hcc2}
\end{eqnarray}
where $H_{\rm cc}^{+-} = \left( H_{\rm cc}^{-+} \right)^{\dag}$. For the matrix elements of ${\bm k} \cdot {\bm p}$ one has
\begin{widetext}
\hspace{-2.0mm}
\begin{eqnarray} \label{Hcc-+}
  H_{\rm cc}^{-+}({\bm k} \cdot {\bm p}) &=&
  \setlength{\arraycolsep}{0.4mm}
  \left(\begin{array}{cccccc}  
P_6k_+&-P_6k_-&Q_2k_+&-Q_2k_-&0&P_5k_z \\ 
P_6k_+&P_6k_-&-Q_2 k_+&-Q_2 k_-&P_5k_z&0 \\ 
-\sqrt{2}P_6k_-&0&0&-P_5k_z&-Q_2k_+&-Q_2k_+\\ 
0&\sqrt{2}P_6k_+&-P_5k_z&0&-Q_2k_-&Q_2k_- \\ 
P_4k_z&0&0&\sqrt{2}P_3k_+&P_3k_-&P_3k_- \\ 
0&P_4k_z&-\sqrt{2}P_3k_-&0&P_3k_+&-P_3k_+ 
\end{array} \right) \\ \label{Hvc-+}
H_{\rm vc}({\bm k} \cdot {\bm p}) &=&
  \setlength{\arraycolsep}{0.4mm}
  \left(\begin{array}{cccccc}  
P_2k_z&0&-Q_1k_-&Q_1k_+&P_1k_+&-P_1k_-\\ 
0&P_2k_z&-Q_1k_-&-Q_1k_+&P_1k_+&P_1k_- \\ 
-Q_1k_+&-Q_1k_+&-P_2k_z&0&\sqrt{2}P_1k_-&0 \\ 
Q_1k_-&-Q_1k_-&0&-P_2k_z&0&-\sqrt{2}P_1k_+
\end{array} \right)
\end{eqnarray}
\end{widetext}
with $k_{\pm}=k_x\pm i k_y$. For the matrix $H_{\rm cc}^{-+}$, we use the ``bra'' and ``ket'' basis functions in the order
$c6\dots c1$ and $c12 \dots c7$ and, for the matrix $H_{\rm vc}^{-+}$,
the ``bra'' basis functions are ordered from $v1$ to $v4$ and
  ``ket'' ones from $c6$ to $c1$. For the operator ${\bm k} \cdot \delta {\bm \pi}$, the matrix elements are as follows
\begin{widetext}
\hspace{-2.0mm}\begin{eqnarray}
  &&H_{\rm cc}({\bm k}\cdot \delta{\bm \pi}) =
  \setlength{\arraycolsep}{0.4mm}
  \left(\begin{array}{cccccc}
\alpha_6k_+ &-\alpha_6k_-&(\beta_2+\alpha_5)k_+&(-\beta_2-\alpha_5)k_-&0&2\sqrt{2}\beta_2k_z\\
\alpha_6k_+ &\alpha_6k_-&(-\beta_2+\alpha_5)k_+&(-\beta_2+\alpha_5)k_-&-2\sqrt{2}\beta_2k_z&0\\
\sqrt{2}\alpha_6k_-&2\sqrt{2}\alpha_6k_z&0&0&(\beta_2+\alpha_5)k_+&(\beta_2-\alpha_5)k_+\\
2\sqrt{2}\alpha_6k_z&-\sqrt{2}\alpha_6k_+&0&0&(\beta_2+\alpha_5)k_-& (-\beta_2+\alpha_5)k_- \\
0&\alpha_4k_+&-2\sqrt{2}\alpha_3k_z&-\sqrt{2}\alpha_3k_+&\alpha_3k_-&\alpha_3k_-\\
-\alpha_4k_-&0&\sqrt{2}\alpha_3k_-&-2\sqrt{2}\alpha_3k_z&\alpha_3k_+&-\alpha_3k_+\\
  \end{array} \right)\:, 
\label{Hcc} \\
   && H_{\rm vc} ({\bm k}\cdot \delta{\bm \pi})=
    \setlength{\arraycolsep}{0.4mm}
    \left(\begin{array}{cccccc}
2\sqrt{2}\beta_1k_z&0&(\beta_1+\alpha_2)k_-&(-\beta_1-\alpha_2)k_+&\alpha_1 k_+&-\alpha_1 k_-\\
0&-2\sqrt{2}\beta_1k_z&(\beta_1-\alpha_2)k_-&(\beta_1-\alpha_2)k_+&\alpha_1 k_+&\alpha_1 k_-\\
(-\beta_1+\alpha_2)k_+&(-\beta_1-\alpha_2)k_+&0&0&-\sqrt{2}\alpha_1k_-&-2\sqrt{2}\alpha_1k_z\\
(\beta_1-\alpha_2)k_-&(-\beta_1-\alpha_2)k_-&0&0&-2\sqrt{2}\alpha_1k_z&\sqrt{2}\alpha_1k_+
    \end{array}\right)\:.
\label{Hvc}
  \end{eqnarray}
\end{widetext}
The coefficients $P_1, P_2 \dots$ and $\alpha_1, \alpha_2\dots$ are purely
imaginary and $Q_1$, $Q_2$, $\beta_1$, $\beta_2$ are real and
  are also listed in Table~\ref{table1}. To illustrate
the applied method to calculate the matrices (\ref{Hcc-+})--(\ref{Hvc}), we
consider the matrix elements between the $L$-point Bloch functions
$\Psi_{j'} = L^-_4, L^-_5, L^-_{6(1)}, 
L^-_{6(2)}$ and $\Phi_j = L^+_4, L^+_5, L^+_{6(1)},L^+_{6(2)}$, respectively for
 $j', j = 1,2,3,4$, formed from the single-group states $L_{3'}$ and
$L_3$ (See Table~\ref{tableX} in Appendix~\ref{appA}). 
Choosing the particular basis states $\left \vert {L_{3'}, i'} \rangle \right.$ ($i' = 1,2$) and 
$\left \vert {L_{3}, i} \rangle \right.$ ($i = 1,2$) we calculate, by using the symmetry considerations, the matrix elements of operators ${\bm k} \cdot {\bm p}$ and $U_{\alpha} = (\hbar^2/2 m_e c^2) 
\left[ {\bm \nabla} V({\bm r}) \times {\bm k} \right]_{\alpha}$ between them.
Let us denote these matrix elements by $M_{ij}$ and $U_{\alpha; ij}$. Now if we present the states $\Psi_{j'}$ and $\Phi_j$ as linear combinations $\sum_{i's'} C_{j', i' s'} \alpha_{s'} \left\vert {L_{3'}, i'} \rangle \right.$ and $\sum_{is} C_{j, i s} \alpha_s \left\vert {L_{3}, i} \rangle \right.$, where $\alpha_s$ is the spin-up state $\uparrow$ for $s=1/2$ and spin-down state $\downarrow$ for $s=-1/2$, then the matrix elements between the spinor states are given by
\begin{eqnarray}
\langle \Psi_{j'} | {\bm k} \cdot {\bm p} | \Phi_j \rangle &=& \sum\limits_{i'i s} C_{j', i' s}^* M_{i'i} C_{j, i s}\:, \nonumber\\
\langle \Psi_{j'} | {\bm k} \cdot \delta {\bm \pi} | \Phi_j \rangle &=& 
\sum\limits_{i'i s' s} C_{j', i' s'}^* ({\bm U}_{i'i} \cdot {\bm \sigma}_{s's}) C_{j, i s}\:. \nonumber
\end{eqnarray}

\begin{figure}[htb]
  \begin{center}
    \includegraphics[width=10.0cm]{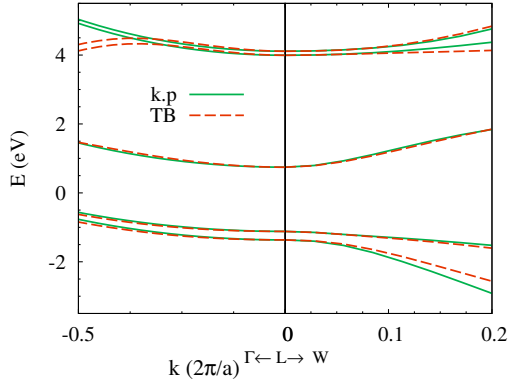}
  \end{center}
  \caption{(Color online) The lowest conduction bands and highest
    valence bands {\em vs.} wave vector near the $(\pi/a)(1,1,1)$ point. 
The origin of  wave vector (the zero point in the figure) 
is taken to be at  the $(\pi/a)(1,1,1)$ point. 
The green solid curves represent the result calculated via the
${\bm k}\cdot{\bm p}$ model and the red dashed curves describe the one
    obtained via the  $sp^3d^5s^{\ast}$ tight-binding (TB) model.\cite{Jancu1}}
  \label{fig1}
\end{figure}

The zero point of wave vector is located at
$(\pi/a)(1,1,1)$ point. $a=5.66$\,\AA\;is the lattice constant
of Ge.\cite{PeterYu1}   One can see that the ${\bm k}\cdot{\bm p}$
and $(\nabla V\times {\bf k})\cdot{\bm \sigma}$ terms couple the bases
with different parities and the $(\nabla V\times {\bf p})\cdot{\bm
  \sigma}$ terms connect bases with the same parity. We determine
the ${\bm k}\cdot{\bm p}$ parameters (given in Table~\ref{table1}) 
by fitting with an $sp^3d^5s^{\ast}$ tight-binding
model.\cite{Jancu1} One can see from Fig.~\ref{fig1} that the lowest three
conduction bands and two highest valence bands calculated from the
${\bm k}\cdot{\bm p}$ parameters 
fit well with the tight-binding results.

\begin{table}[htb]
  \caption{The parameters in the $16\times16$ ${\bf k}\cdot{\bf p}$
    Hamiltonian in the vicinity of the $L$ point. These
      parameters are obtained by fitting with the band structure in an $sp^3d^5s^{\ast}$ tight-binding
model.\cite{Jancu1}}
  \label{table1} 
  \begin{tabular}{ll|ll}
    \hline
    \hline
&&&\\[-6pt]
    &\;  eV$\cdot$ nm\;&&\quad\quad  eV\quad\quad  \\ [4pt]  
    \hline
&&&\\[-6pt]
    $Q_1$&\;\;\;$0.52$&\;$E_{v1}$&\quad$-1.119$ \\[4pt]
    $Q_2$&\;\;\;$0.26$ &\;$E_{v2}$&\quad$-1.365$ \\[4pt]
    $P_1$&\;\;\;$0.52\, i$&\;$E_{c1}$&\quad\;\;\;$0.747$\\[4pt]
    $P_2$&\;\;\;$0.32\, i$&\;$E_{c2}$&\quad\;\;\;$3.990$ \\[4pt]
    $P_3$&\;\;\;$0.31\, i$&\;$E_{c3}$&\quad\;\;\;$4.110$ \\[4pt]
    $P_4$&\;\;\;$0.37\, i$&\;$E_{c4}$&\quad\;\;\;$8.349$\\[4pt]
    $P_5$&$-0.35\, i$&\;$E_{c5}$&\quad\;\;\;$8.354$\\[4pt]
    $P_6$&$-0.20\, i$ &\;$E_{c6}$&\quad\;\;\;$9.109$\\[4pt]
    $\beta_{1}$&\;\;\;$0.03$&\;$\Delta_1$&\quad\;\;\;$0.012$\\[4pt]
    $\beta_{2}$&\;\;\;$0.01$&\;$\Delta_{2}$&\quad\;\;\;$0.089$\\[4pt]
    $\alpha_{1}$&$-0.01\, i$&\;$\Delta_{3}$&\quad\;\;\;$0.041$\\[4pt]
    $\alpha_2$&\;\;\;$0.02\, i$&\;$\Delta_{4}$&\quad\;\;\;$0.097$\\[4pt]
    $\alpha_{3}$&\;\;\;$0.02\, i$&&\\[4pt]
    $\alpha_{4}$&\;\;\;$0.04\, i$&&\\[4pt]
    $\alpha_{5}$&\;\;\;$0.04\, i$&&\\[4pt]
    $\alpha_{6}$&\;\;\;$0.02\, i$&&\\[4pt]
    \hline
    \hline
\end{tabular}
\end{table}

For the other three $L$ points, i.e., $(\pi/a)(-1,-1,1)$,
$(\pi/a)(-1,1,1)$ and
$(\pi/a)(1,-1,1)$, the basis functions and
Hamiltonian share the same forms with those in Appendix~\ref{appA} and  
Eqs.\,(\ref{Htotal})-(\ref{Hvc}), while the coordinate
system varies.

It is noted that a ${\bm k}\cdot{\bm p}$
matrix  larger than or equal to
$12\times12$ is necessary to fit the structure
of the lowest 
conduction and highest valence bands 
without the remote-band influence. These bands are
 pertinent to the electron spin relaxation. Via the L\"owdin
partitioning,\cite{Winkler1} the effective masses 
of $i$-th band can be expressed as 
\begin{subequations}
\begin{align}
  \frac{m_{\rm e}}{m^{i}_{z}}&=1+\frac{2m_{\rm
      e}}{\hbar^2}\sum_{j\not=i}\fr{|H_{ij}|_{k_x,k_y=0}^2}{(E_i-E_j)k^2_z},\\
  \frac{m_{\rm e}}{m^{i}_{x(y)}}&=1+\frac{2m_{\rm
      e}}{\hbar^2}\sum_{j\not=i}\fr{|H_{ij}|_{k_z,k_y(k_x)=0}^2}{(E_i-E_j)k^2_{x(y)}}.
\end{align}
\end{subequations}
  For the lowest conduction band, $m_z=1.36\,{m}_{\rm e}$, which
confirms the importance of including the fourth conduction
band.\cite{Jancu1}

Moreover, with our $16\times16$ ${\bm k}\cdot{\bm p}$
 Hamiltonian, it is easy to obtain the
eigenstates of lowest conduction bands at the $L$ points,

\begin{subequations}\label{eigenL}
\begin{align}
  \varphi_{\frac{1}{2}}&=\frac{1}{A}[L^+_{6(1)}(L_1)-cL^+_{6(1)}(L_3)]\nonumber\\
  &=
  \frac{1}{A}[L_1\up-c(L_{3y}- i L_{3x})\down],\\
  \varphi_{-\frac{1}{2}}&=\frac{1}{A}[L^+_{6(2)}(L_1)-cL^+_{6(2)}(L_3)]\nonumber\\
  &=\frac{1}{A}[L_1\down+c(L_{3y}+ i
  L_{3x})\up],
\end{align}
\end{subequations}
where $L_{3x}\sim -zy$, $L_{3y}\sim zx$ and
$\up$ ($\down$) is the spin up (down)
eigenstate along the $z$ direction in the corresponding coordinate system. $A$ is the renormalization
coefficient and $c$ is a real parameter which can be calculated
  directly by diagonalizing the ${\bm k}\cdot{\bm p}$ Hamiltonian. 
Clearly the conduction electron states
are spin-mixing at the $L$ points. In this work the $L_1$, $L_{3x}$
and $L_{3y}$ are taken to be purely real. 

\section{ELECTRON-PHONON SCATTERING} 
We investigate the
electron-phonon scatterings of $L$ valleys and $\Gamma$ valley in Ge,
{i.e.}, the intra-$\Gamma$/$L$ valley, inter--$\Gamma$-$L$ valley and inter-$L$
  valley scatterings. These scatterings  are fundamental to understand
  the spin dynamics in the optical orientation of spin and the hot-electron relaxation.\cite{Tang-Ge1,Loren-Ge2,Sipe-Optical1,Pzhang1} The
  complete scattering matrices and selection rules are
  determined with the invariant method and subgroup technique. Also,
 we analyze the Elliott\cite{Elliott1}
  and Yafet\cite{Yafet1} mechanisms in these electron-phonon scatterings
  facilitated by the ${\bm k}\cdot{\bm p}$ basis functions (See
    Table~\ref{tableX} in Appendix~\ref{appA}) and Hamiltonian  obtained
around the $L$ point.\cite{Voon1,Lax-Selection1,Lax2} The coefficients in the scattering
matrices are obtained by the pseudo-potential method, which further confirms our selection rules and wave-vector dependence. 

We first derive the time-reversal constraint
on the wave-vector dependence of scattering matrix. The spin-related
scattering matrix in the centrosymmetric crystal  can be
generally written as 
\begin{equation}
\hat{M}_{{\bf k},{\bf k}^{\prime}}=A_{{\bf k},{\bf k}^{\prime}}\hat{I}+{\bf B}_{{\bf k},{\bf k}^{\prime}}\cdot{\bm \sigma},\label{matrix}
\end{equation}
where {\bf k}$^{\prime}$ ({\bf k}) is the wave
vector of initial (final)
  electronic state and the spin eigenstates are along $\hat{z}$
  direction. $\hat{I}$ is the $2\times2$ identity matrix and ${\bm
    \sigma}$ are the Pauli matrices.  Via the time-reversal operator, it's easy to
obtain the time-reversal constraint on the wave-vector dependence
\begin{subequations}\label{time_reversal}
\begin{align}
A_{{\bf k},{\bf k}^{\prime}}&=A^{\ast}_{{\bf k}^{\prime},{\bf k}}=A_{-{\bf k}^{\prime},-{\bf k}},\\
{\bf B}_{{\bf k},{\bf k}^{\prime}}&={\bf B}^{\ast}_{{\bf k}^{\prime},{\bf k}}=-{\bf B}_{-{\bf k}^{\prime},-{\bf k}},
\end{align}
\end{subequations} 
where $A_{{\bf k},{\bf k}^{\prime}}$ (${\bf B}_{{\bf k},{\bf k}^{\prime}}$) is purely
real (imaginary).

Moreover, the spin-orientation dependence of the electron-phonon
  scattering can be obtained easily with this scattering matrix. The scattering elements  with spin
eigenstates along  an arbitrary
direction
$\hat{n}=(\sin{\theta}\cos{\phi},\sin{\theta}\sin{\phi},\cos{\theta})$
can be expressed as
\begin{subequations}\label{orientation}
\begin{align}
M_{{\bf k},{\bf k}^{\prime};\up,\up}=&A_{{\bf k},{\bf k}^{\prime}}+\cos{\theta}B_{{\bf k},{\bf k}^{\prime};z}\nonumber\\&\mbox{}+\sin{\theta}(\cos{\phi}B_{{\bf k},{\bf k}^{\prime};x}+\sin{\phi}B_{{\bf k},{\bf k}^{\prime};y}),\label{orientationA}\\
M_{{\bf k},{\bf k}^{\prime};\up,\down}=&-\sin{\theta}B_{{\bf k},{\bf k}^{\prime};z}+(\cos{\theta}\cos{\phi}+ i\sin{\phi})B_{{\bf k},{\bf k}^{\prime};x}\nonumber\\&\mbox{}+(\cos{\theta}\sin{\phi}- i\cos{\phi})B_{{\bf k},{\bf k}^{\prime};y}.
 \label{orientationB}
\end{align}
\end{subequations}

We further analyze the wave-vector order of
  scattering matrix element, which is given by\cite{Tang-Ge1,Dery_Si1}
\begin{eqnarray}
&\sqrt{\fr{\hbar}{2\rho
      V\omega_{{\bf q}}}}\sqrt{n_{{\bf q}}+\fr{1}{2}\pm\fr{1}{2}}M_{{\bf k},{\bf k}^{\prime};s,s^{\prime}}\label{matrix_element}\nonumber\\
&=\lan{\bf k},s,n_{{\bf q}}|H_{\rm
  ep}|{\bf k}^{\prime},s^{\prime},n_{{\bf q}}\pm1\ran,
\end{eqnarray} 
with
  $\rho$, $V$, $\omega_{{\bf q}}$ and $n_{{\bf q}}$ representing the crystal
  mass density, crystal volume, phonon frequency and phonon occupation,
  respectively. The matrix element can be expressed in order of
$\delta{\bf q}=({\bf k}^{\prime}-{\bf k}_{\rm i})-({\bf k}-{\bf k}_{\rm f})$,\cite{Tang-Ge1,Dery_Si1}
\begin{equation}
M_{{\bf k},{\bf k}^{\prime};s,s^{\prime}}=D_{0,s,s^{\prime}}+{\bf D}_{1,s,s^{\prime}}\cdot{\delta}{\bf
  q}+\delta{\bf q}\cdot{\bf D}_{2,s,s^{\prime}}\cdot\delta{\bf q}+...\\
\end{equation}
with ${\bf k}_{\rm i}$ (${\bf k}_{\rm f}$) being the valley center of the initial
(final) state and $\delta{\bf q}\ll 2\pi/a$.\cite{Ferry1}  There are vanishing
(non-vanishing) zeroth-order contributions to the spin-flip scattering in the intra- (inter-) valley
  scatterings. Hereafter we derive the lowest-order wave-vector dependence of
  the scattering matrix for each phonon mode only due to its dominant
  contribution compared to the corresponding higher order terms.  We
also give the non-zeroth order
contributions to the spin-flip intervalley
electron-phonon scatterings in the spherical-band-approximation for
completeness.

\begin{table}[htb]
\tabcolsep=0.1cm
\renewcommand{\arraystretch}{2.0}
\caption{\label{long_wavelength}
Phonon polarization vectors in the long-wavelength limit ($\mathbf{q}
\ll 2\pi/a$). Here the superscript ``$+$'' (``$-$'') represents the
in-phase (out-of-phase). $F_1$ and $F_2$ are the corresponding
normalization coefficients. $G({\bf
  q})=q^4_x(q^2_y-q^2_z)+q^4_y(q^2_z-q^2_x)+q^4_z(q^2_x-q^2_y)$ is one
basis function of $\Gamma^+_2$ irreducible representation in $O_h$
group.\cite{Group1} We have used the elastic continuum approximation for
diamond crystal structures.  Two TA or
TO polarizations can be linearly combined into any other orthonormal ones.
 }
\begin{tabular}{cc}
\hline \hline
$\bf\xi^+_{\rm{TA_1}}, \bf\xi^-_{\rm{TO_1}}(\mathbf{q})$  & $
\frac{1}{F_1}\{q_x[q^2_y(q^2_x-q^2_y)-q^2_z(q^2_z-q^2_x)]$,\\
&\quad\;\;$q_y[q^2_z(q^2_y-q^2_z)-q^2_x(q^2_x-q^2_y)]$,\\&\quad\;\;\;$q_z[q^2_x(q^2_z-q^2_x)-q^2_y(q^2_y-q^2_z)]\} $ \\
$\bf\xi^+_{\rm{TA_2}}, \bf\xi^-_{\rm{TO_2}}(\mathbf{q})$  &
$\frac{G({\bf q})}{F_2}[{\rm sgn}(q_x)|q_yq_z|(q^2_y-q^2_z)$,\\&\quad\quad\; ${\rm
    sgn}(q_y)|q_xq_z|(q^2_z-q^2_x)$,\\&\quad\quad\; ${\rm sgn}(q_z)|q_xq_y|(q^2_x-q^2_y)] $ \\
$\bf\xi^+_{\rm{LA}}, \bf\xi^-_{\rm{LO}}(\mathbf{q})$  &
$\mathbf{q}/|\mathbf{q}|$\\
\hline \hline
\end{tabular}
\end{table}

\subsection{Numerical  method} 
To obtain the coefficients of these scattering matrices, we derive the electron-phonon interaction and
evaluate the matrix elements under an
empirical pseudo-potential model by following
 Ref.~[\onlinecite{Phys.Rev.B_83_165211_2011_Cheng}]. This method has been
successfully used in the calculation of the spin
relaxation time,\cite{Phys.Rev.Lett._104_016601_2010_Cheng} the degree of circular polarization of the
luminescence across the indirect band\cite{Phys.Rev.Lett._105_037204_2010_Li} and indirect
optical injections\cite{Phys.Rev.B_83_165211_2011_Cheng} in bulk
Si, and obtains good agreements with experiments.  In this model,
  the real single particle potential $V(\bm r)=\sum_{i\alpha}v(\bm r-\bm
  R_{i\alpha})$ is replaced by a smooth pseudo potential
  $\tilde{v}(\bm r)=v_L(\bm r)+v_{NL}(\bm r) + v_{so}(r)\bm
    l\cdot\bm \sigma$ which includes the local potential $v_L$,
  the non-local one $v_{NL}$, and the spin-orbit coupling part
  $v_{so}$ with $\bm l$ being the orbital momentum operator. Here the
  subscripts $i$ and $\alpha$ are the indices for the primitive cells of
  the crystal  and atoms in a primitive cell separately.
   This
  pseudo potential is chosen to produce the same single particle energy
  as the real potential  but a much smooth wave function around the 
  nuclei. Therefore, the choice of the pseudo potential is not
  unique. In our calculation, it is taken from
  Ref.~[\onlinecite{Phys.Rev.B_48_14276_1993_Rieger}]. The calculated
electron energies at the band edges of conduction band for
  $\Gamma$ and $L$ valleys match those in the $sp^3d^5s^{\ast}$ tight-binding
model.\cite{Jancu1} By shifting the
atom position $\bm R_{i\alpha}$ by $\bm u_{i\alpha}$ and expanding
$V(\bm r)$ with respect to $\bm u_{i\alpha}$, the linear term in the
expansion is the electron-phonon interaction $H_{ep}$. The atom
displacement is related to phonon operators by $\bm u_{i\alpha}=\sum_{\bm
  q\lambda}(\hbar/\rho\omega_{\bm q})^{1/2}(a_{\bm
  q\lambda}+a_{-\bm q\lambda}^{\dag})\bm\epsilon_{\bm q\lambda}^\alpha
e^{i\bm q\cdot\bm R_i}$, where ${\bm q}$/$\lambda$ represents the phonon
wave vector/mode. We calculate the phonon polarization vectors
$\bm\epsilon_{\bm q\lambda}$ by an adiabatic bond charge
model.\cite{Phys.Rev.B_15_4789_1977_Weber} The calculated energies 
for the phonons involved in the scattering channels studied here are $10.2$, $28.6$ and $33.3$~meV for
$X_3$, $X_1$ and $X_4$ phonons, and $7.4$, $25.6$,
$29.3$ and $35.80$~meV for $L_3$, $L_{2'}$, $L_1$ and
$L_{3'}$ phonons, respectively.

With the pseudo-potential method, we obtain the electron-phonon matrix
elements for the intra-$\Gamma$/$L$ valley, inter--$\Gamma$-$L$ valley  and inter-$L$ valley scatterings
in Ge, and confirm  the selection rules and scattering matrices
  in all cases. The coefficients
  in these scattering matrices are determined by fitting with the
  pseudo-potential results and  are listed in Tables~\ref{table_Gamma}-\ref{table_T_L}. One can see that generally the scattering matrix element
for the spin-conserving process scattering
is much larger (more than 50 times) than that for the spin-flip one.

\subsection{Intravalley scattering}
We first address the intra-$\Gamma$/$L$ valley electron-phonon scatterings
in Ge with the theory of invariants, where the LA, TA and OP
contributions are all included.\cite{Voon1,Bir1}  The contributions of
two TA (three OP) phonon modes are summed up as
their sound velocities (phonon energies) near the
$\Gamma$ point are close to each other.\cite{Dery_Si1} It should be
noted  that, beyond the symmetry of each phonon branch, the wave-vector
dependence of polarization vectors for AC (belong to $\Gamma^-_{15}$
irreducible representation) and OP
(belong to $\Gamma^+_{25}$ irreducible representation) phonon modes at the $\Gamma$ point are well-known and
compact (See Table~\ref{long_wavelength}), which facilitate our
derivation of the wave-vector dependence of the matrix elements. In Ge, the wave-vector-order analysis of the spin-flip process
  has already been performed based on the time-reversal and space-inversion
  symmetry, where the leading-order terms for the AC phonons are
  third-order ($K_lq_mq_n$) and those for the OPs are second-order
  ($K_lq_m$) ($l,m,n\in\{x,y,z\}$).\cite{Yafet1,Dery_Si1,Li-Ge2}

\subsubsection{Intra-$\Gamma$ valley scattering}

In intra-$\Gamma$ scattering case, we apply the conventional
coordinate system for simplify. The $\hat{M}_{{\bf k},{\bf k}^{\prime}}$ should be an
invariant in $O_h$ point group.\cite{Group1,PeterYu1} 
First, one can see that the
zeroth-order contribution of scattering matrix vanishes as
\begin{equation}
\Gamma^-_7\otimes \Gamma^-_7=\Gamma^+_1+\Gamma^+_{15}.
\end{equation} 

\begin{table}[htb]
   \caption{The nonvanishing coefficients in the intra-$\Gamma$ valley
    electron-phonon scattering matrix. The first two columns are the
    coefficients for the AC phonons, and the last two columns show the parameters in
    the OP case. $\Xi_1=8.42$ eV is the coefficient for the LA
    contribution to the  spin-conserving
    scattering.}
  \label{table_Gamma}
  \begin{tabular}{c|c|c}
    \hline    \hline
    &&      \\[-6pt]
    &\;AC,\;eV$\cdot{\rm nm}^2$&\;OP,\;eV$\cdot$ nm \\ [2pt]  
    \hline
    &&      \\[-6pt]
    $R_1$\;&$-0.61$ &  \\[4pt]
    $R_2$\;&\;\;\;$0.55$  &  $4.35$ \\ [4pt]
    $R_3$\;&\;\;\;$0.60$  &  \\[4pt]
    $R_4$\;&$-0.83$  &    \\[4pt]
    $R_5$\;&$-1.19$ &  $0.0055$   \\[4pt]
    $R_6$\;&\;\;\;$ 2.00$  & $0.35$\\[4pt]
    \hline
    \hline
  \end{tabular}
\end{table} 
For  the $O_h$  point group, the symmetries of spin-dependent matrices
  are given by
\begin{equation}
\hat{I}\sim
  \Gamma^+_1,\quad\quad {\bm \sigma}\sim
  \Gamma^+_{15}.
\end{equation} 
Therefore, from the invariance of the scattering matrix $\hat{M}_{{\bf k},{\bf k}^{\prime}}$, the symmetries of $A_{{\bf k}, {\bf k}^{\prime}}$ and ${\bf B}_{{\bf k},
  {\bf k}^{\prime}}$ can be determined by
\begin{equation}
A_{{\bf k}, {\bf k}^{\prime}}\sim
  \Gamma^+_1,\quad\quad {\bf B}_{{\bf k},{\bf k}^{\prime}}\sim
  \Gamma^+_{15}.
\end{equation} 
Then we can construct their explicit forms, which are
  functions of the wave vectors and phonon polarization vectors. It
  should be noted that the
time-reversal constraint [Eq.\,(\ref{time_reversal})] must be
fulfilled. The  symmetry-allowed terms are given by
\begin{eqnarray} \label{bkk'}
A_{{\bf k}, {\bf k}^{\prime}}&=&\Xi_1(u_{xx}+u_{yy}+u_{zz})
\label{G_A}\\
{\bf B}_{{\bf k}, {\bf k}^{\prime}}& = &i \sum\limits_{j=1}^6 R_j {\bf S}_{{\bf k}, {\bf k}^{\prime}}^{(j)},\label{G_B}
\end{eqnarray}
with
\begin{subequations}\label{G_Sz}
\begin{align}
S^{(1)}_z &= ({\bf K} \times {\bf q} )_z( u_{xx} + u_{yy} + u_{zz} ), \label{G1} \\
S^{(2)}_z &= ({\bf K} \times {\bf q} )_x u_{zx} + ({\bf K}\times{\bf q} )_y u_{yz}, \label{G2}\\
S^{(3)}_z &= ({\bf K} \times {\bf q} )_z u_{zz},\label{G3} \\
S_z^{(4)} &= (K_x q_y + K_y q_x) (u_{xx} - u_{yy}),\label{G4} \\
S_z^{(5)} &= (K_y q_z + K_z q_y) u_{zx} - (K_z q_x + K_x q_z) u_{yz}, \label{G5} \\
S_z^{(6)} &= (K_x q_x - K_y q_y) u_{xy}\label{G6}.
\end{align}
\end{subequations}
Here we omit the subscripts ${\bf k}, {\bf k}^{\prime}$
 for simplicity in Eqs.\;(\ref{G1})-(\ref{G6}).
The components of ${\bf S}_{{\bf k},{\bf k}^{\prime}}$ are connected by the coordinate
permutation.  ${\bf K}={\bf k}^{\prime}+{\bf k}$ and ${\bf q}={\bf k}^{\prime}-{\bf
  k}=\delta{\bf q}$. ${\bf
  K}$ and ${\bf q}$ are in the same order.  For AC modes,
$u_{\alpha\beta}=\frac{1}{2}(q_{\alpha}\xi^+_{\rm
  ac,\beta}+q_{\beta}\xi^+_{\rm ac,\alpha})$ with ${\rm \alpha}$ $({\beta})\in\{x,y,z\}$.  In OP modes,  we set $u_{\alpha \alpha} = 0$ and replace $u_{yz}, u_{zx},
u_{xy}$ by optical vibration amplitudes $\xi^-_{{\rm op},x},\xi^-_{{\rm
    op},y},\xi^-_{{\rm op},z}$. Here ${\bm \xi}^+_{\rm ac}$ and
${\bm \xi}^-_{\rm op}$ are the phonon polarization vectors in the
long-wavelength limit, which are given in
Table~\ref{long_wavelength}. The coefficients $\Xi_i$ and $R_i$ are
listed in Table~\ref{table_Gamma}.

Finally, from the numerical pseudo-potential calculation, we can see
that for the spin-conserving scattering, the scattering element for
LA phonon branch is 
 more than 2 orders of magnitude larger than that for other modes, as the
 lowest order spin-conserving element only exists in LA case 
[See Eq.\;(\ref{G_A})].  While for
 the spin-flip scattering, the OP contribution is about 4
 orders of magnitude larger than the AC contribution.

\subsubsection{Intra-$L$ valley scattering}

Recently, Tang {\em et al.}\cite{Tang-Ge1} and Li {\em et
  al.}\cite{Li-Ge2} investigated the  intra-$L$ electron-phonon
scattering in Ge both analytically and numerically. Tang {\em et
  al.}\cite{Tang-Ge1}  gave the average absolute values of the
scattering elements for  the
AC contribution, and for the spin-conserving process
of the OP contribution in the spherical-band-approximation.  Li {\em et al.}\cite{Li-Ge2} derived the approximated wave-vector
dependence for the AC contribution, where the contributions of the three
AC phonon modes were summed up.  Here we demonstrate the complete
and detailed wave-vector dependence for the intra-$L$ valley scattering matrix via
the theory of invariants, where the LA, TA and OP contributions
  are all considered.

In the intra-$L$ valley scattering, the scattering matrix
$\hat{M}_{{\bf k},{\bf k}^{\prime}}$ should be an invariant in the $D_{3d}$ point
group.\cite{Group1,PeterYu1} 
We consider the $(\pi/a)(1,1,1)$ point and choose the [111]
  coordinate system for simplicity. The results in other three $L$
  points can be obtained by coordinate rotation. Here ${\bf
k}_{\rm i}={\bf
k}_{\rm f}=(\pi/a)(1,1,1)$ and should be extracted from the initial
and final wave-vectors. In this
case,  the zeroth-order contribution of the OP modes to
spin-conserving scattering  exists,
\begin{equation}
  L^+_6\otimes L^+_6=\Gamma^+_1+\Gamma^+_{25},\label{intraL^{+}_selection}
\end{equation}
and the zeroth-order spin-flip elements are forbidden by time-reversal symmetry.\cite{Tang-Ge1}

For $D_{3d}$ point group, the symmetries of the
  spin-dependent matrices are reduced to
\begin{equation}
\hat{I}\sim
  L_1,\quad\quad \sigma_z\sim
  L_{2},\quad\quad  (\sigma_x,\sigma_y)\sim
  (L_{3x},L_{3y}).
\end{equation}
To ensure the invariance of $\hat{M}_{{\bf
      k},{\bf k}^{\prime}}$, the symmetries of the $A_{{\bf k},{\bf k}^{\prime}}$ and ${\bf B}_{{\bf k},{\bf
    k}^{\prime}}$ are given by
\begin{eqnarray}
&&A_{{\bf k},{\bf k}^{\prime}}\sim
  L_1,\quad\quad B_{{\bf k},{\bf k}^{\prime};z}\sim
  L_{2},\quad\quad\\ &&(B_{{\bf k},{\bf k}^{\prime};x},B_{{\bf k},{\bf k}^{\prime};y})\sim
  (L_{3x},L_{3y}).
\end{eqnarray}
Then, the explicit wave-vector of the scattering matrix can be constructed as
well, which is much more complex than that in intra-$\Gamma$ case
as the symmetry is reduced.
\begin{eqnarray}\label{L^{+}_A}
  A_{{\bf k},{\bf k}^{\prime}}&=&\Xi_1(u_{xx}+u_{yy})+\Xi_2u_{zz},\\
  B_{{\bf k},{\bf k}^{\prime};z}&=&i\sum\limits_{j=1}^8
C_j S^{(j)}_{{\bf k},{\bf k}^{\prime};z},\\
  (B_{{\bf k},{\bf k}^{\prime};x},B_{{\bf k},{\bf
      k}^{\prime};y})&=&i\sum\limits_{j=1}^{18}
R_j (S^{(j)}_{{\bf k},{\bf k}^{\prime};x},S^{(j)}_{{\bf k},{\bf k}^{\prime};y}),
\end{eqnarray}
 with
\begin{subequations}\label{L^{+}_Bz}
  \begin{align}
    S^{(1)}_{z}&=(u_{xx}+u_{yy})({\bf K}\times{\bf q})_z,\label{Lz1}\\
    S^{(2)}_{z}&=u_{zz}({\bf K}\times{\bf q})_z,\label{Lz2}\\
    S^{(3)}_{z}&=2u_{xy}(q_xK_x-q_yK_y)\nonumber\\&\quad-(u_{xx}-u_{yy})(q_xK_y+q_yK_x),\label{Lz3}\\
    S^{(4)}_{z}&=q_z[(u_{xx}-u_{yy})K_x-2u_{xy}K_y],\label{Lz4}\\
    S^{(5)}_{z}&=K_z[(u_{xx}-u_{yy})q_x-2u_{xy}q_y],\label{Lz5}\\
    S^{(6)}_{z}&=u_{yz}(K_xq_y+K_yq_x)-u_{xz}(K_xq_x-K_yq_y),\label{Lz6}\\
    S^{(7)}_{z}&=q_z(u_{xz}K_y-u_{yz}K_x),\label{Lz7}\\
    S^{(8)}_{z}&=K_z(u_{xz}q_y-u_{yz}q_x),\label{Lz8}
  \end{align}
\end{subequations}
and
\begin{subequations}\label{L^{+}_Bxy}
  \begin{align}
    (S^{(1)}_{x},S^{(1)}_{y}) &=(u_{xx}+u_{yy})q_z(-K_y,K_x),\label{Lx1}\\
    (S^{(2)}_{x},S^{(2)}_{y}) &=u_{zz}q_z(-K_y,K_x),\label{Lx2}\\
    (S^{(3)}_{x},S^{(3)}_{y}) &=(u_{xx}+u_{yy})K_z(-q_y,q_x),\label{Lx3}\\
    (S^{(4)}_{x},S^{(4)}_{y}) &=u_{zz}K_z(-q_y,q_x),\label{Lx4}\\
    (S^{(5)}_{x},S^{(5)}_{y})
    &=(u_{xx}+u_{yy})(q_xK_x-q_yK_y,\nonumber\\&\mbox{}\quad-q_yK_x-q_xK_y),\label{Lx5}\\
    (S^{(6)}_{x},S^{(6)}_{y})
    &=u_{zz}(q_xK_x-q_yK_y,-q_yK_x-q_xK_y),\label{Lx6}\\
    (S^{(7)}_{x},S^{(7)}_{y}) &=q_zK_z(u_{xx}-u_{yy},-2u_{xy}),\label{Lx7}\\
    (S^{(8)}_{x},S^{(8)}_{y}) &=(q_xK_x+q_yK_y)(u_{xx}-u_{yy},-2u_{xy}),\label{Lx8}\\
    (S^{(9)}_{x},S^{(9)}_{y}) &=q_zK_z(-u_{yz},u_{xz}),\label{Lx9}\\
    (S^{(10)}_{x},S^{(10)}_{y}) &=(q_xK_x+q_yK_y)(-u_{yz},u_{xz}),\label{Lx10}\\
    (S^{(11)}_{x},S^{(11)}_{y}) &=[(u_{xx}-u_{yy})(q_xK_x-q_yK_y)\nonumber\\&\mbox{}\quad\mbox{}-2u_{xy}(q_xK_y+q_yK_x),\nonumber\\&\quad(u_{xx}-u_{yy})(q_xK_y+q_yK_x)\nonumber\\&\mbox{}\quad\mbox{}+2u_{xy}(q_xK_x-q_yK_y)],\label{Lx11}\\
    (S^{(12)}_{x},S^{(12)}_{y})
    &=[u_{xz}(q_xK_y+q_yK_x)\nonumber\\&\quad\mbox{} -u_{yz}(q_xK_x-q_yK_y),\nonumber\\&\quad -u_{xz}(q_xK_x-q_yK_y)\nonumber\\&\quad\mbox{}-u_{yz}(q_xK_y+q_yK_x)],\label{Lx12}\\
    (S^{(13)}_{x},S^{(13)}_{y}) &=K_z[(u_{xx}-u_{yy})q_y-2u_{xy}q_x,\nonumber\\&\quad(u_{xx}-u_{yy})q_x+2u_{xy}q_y],\label{Lx13}\\
    (S^{(14)}_{x},S^{(14)}_{y})
    &=q_z[(u_{xx}-u_{yy})K_y-2u_{xy}K_x,\nonumber\\&\quad(u_{xx}-u_{yy})K_x+2u_{xy}K_y],\label{Lx14}\\
    (S^{(15)}_{x},S^{(15)}_{y}) &=K_z(u_{yz}q_y-u_{xz}q_x,u_{yz}q_x+u_{xz}q_y),\label{Lx15}\\
    (S^{(16)}_{x},S^{(16)}_{y})
    &=q_z(u_{yz}K_y-u_{xz}K_x,u_{yz}K_x+u_{xz}K_y),\label{Lx16}\\
    (S^{(17)}_{x},S^{(17)}_{y}) &=({\bf K}\times
    {\bf q})_z(2u_{xy},u_{xx}-u_{yy}),\label{Lx17}\\
    (S^{(18)}_{x},S^{(18)}_{y})& =({\bf K}\times
    {\bf q})_z(u_{xz},u_{yz}).\label{Lx18}
   \end{align}
\end{subequations}

Here we have again omitted the subscripts ${\bf k},{\bf k}^{\prime}$ in
  Eqs.\;(\ref{Lz1})-(\ref{Lx18}). 
The components of ${\bf S}_{{\bf k},{\bf k}^{\prime}}$ cannot be connected by the
  coordinate permutation now due to the reduced symmetry. Here
  $u_{xx}+u_{yy}$ corresponds  to $\xi^-_{{\rm op},z}$ for OP modes,
  and other expressions are the same as those in the intra-$\Gamma$
  case. It should be noted that the quadratic sum of
OP contributions of $A_{{\bf k},{\bf k}^{\prime}}$ is
constant [See Eq.\;(\ref{L^{+}_A})] and Table~\ref{long_wavelength},
which is in agreement with the selection rule
[Eq.\,(\ref{intraL^{+}_selection})]. Moreover,  Eqs.\,(\ref{Lz1}),
(\ref{Lz2}) and (\ref{Lx3})-(\ref{Lx6}) are in agreement with the
approximated analytical
forms of the intra-$L$ scattering matrix in Li's work.\cite{Li-Ge2} 
Obviously our wave-vector-dependent  scattering matrices are
 more detailed than those in previous works.\cite{Tang-Ge1,Li-Ge2}

Finally the coefficients  $\Xi_i$, $C_i$ and $R_i$  in the intra-$L$ valley
  electron-phonon scattering matrix are calculated from
the pseudo-potential calculation with the non-vanishing ones listed in 
 Table~\ref{table_L}. It is interesting to see that  for 
the spin-conserving scattering, unlike the
  intra-$\Gamma$ case, the LA contribution is close to the TA one, as
the lowest-order spin-conserving elements exist in both cases. For
  the spin-flip scattering, the ratio of
OP contribution and AC contribution is close to that
in intra-$\Gamma$ case. Our AC contribution and the spin-conserving OP
contribution are in the same order as those in previous 
work.\cite{Tang-Ge1,Li-Ge2}

\begin{table}[h]
  \caption{The nonvanishing coefficients in the intra-$L$ valley
    electron-phonon scattering matrix. The first two columns are the
    coefficients for AC phonon modes and the last two columns show that in
    OP case.}
  \label{table_L}
  \begin{tabular}{ll|ll}
    \hline
    \hline
    &&&      \\[-6pt]
    AC&\;\;\;\;eV\;&\;OP&\;\;\;\;eV/nm\; \\ 
    \hline
    &&&      \\[-6pt]
    $\Xi_1$&\;$-8.6$&\;\;$\Xi_1$&\;\;\;\;$43.8$\\
    $\Xi_2$&\;\;\;\;$5.8$&\;&  \\
    \hline
   \hline
    &&&      \\[-6pt]
    AC&\;\;\;\;eV$\cdot{\rm nm}^2$\; &\;OP&\;\;\;\;eV$\cdot{\rm nm}$\;\\ 
    \hline 
    &&&      \\[-6pt]
    $R_1$&\;$-3.5\times10^{-3}$\; &\;\;$R_1$&\;\;\;\;$6.0\times10^{-3}$\; \\
    $R_2$&\;\;\;\;$2.1\times10^{-3}$\; &\;\;$R_3$&\;$-2.0\times10^{-2}$\; \\
    $R_3$&\;\;\;\;$4.2\times10^{-3}$\; &\;\;$R_5$&\;\;\;\;$1.3\times10^{-2}$\; \\
    $R_4$&\;\;\;\;$0.5\times10^{-4}$\; &\;\;$R_9$&\;\;\;\;$6.9\times10^{-3}$\; \\
    $R_5$&\;\;\;\;$2.0\times10^{-3}$\; &\;\;$R_{10}$&\;$-2.8\times10^{-2}$\; \\
    $R_6$&\;\;\;\;$3.5\times10^{-3}$\; &\;\;$R_{12}$&\;$-6.5\times10^{-2}$\; \\
    $R_7$&\;$-1.0\times10^{-5}$\; &\;\;$R_{15}$&\;\;\;\;$2.5\times10^{-2}$\; \\
    $R_8$&\;$-1.2\times10^{-3}$\; &\;\;$R_{16}$&\;$-1.8\times10^{-2}$\; \\
    $R_9$&\;$-5.9\times10^{-4}$\; &\;\;$R_{18}$&\;$-7.0\times10^{-2}$\; \\
    $R_{10}$&\;\;\;\;$1.0\times10^{-2}$\; &\;&\;\; \\
    $R_{11}$&\;$-6.0\times10^{-4}$\; &\;&\;\; \\
    $R_{12}$&\;$-4.9\times10^{-3}$\; &\;&\;\; \\
    $R_{13}$&\;\;\;\;$3.3\times10^{-3}$\; &\;&\;\; \\
    $R_{14}$&\;\;\;\;$5.0\times10^{-3}$\; &\;&\;\; \\
    $R_{15}$&\;\;\;\;$2.0\times10^{-3}$\; &\;&\;\; \\
    $R_{16}$&\;\;\;\;$8.6\times10^{-4}$\; &\;&\;\; \\
    $R_{17}$&\;$-2.5\times10^{-3}$\; &\;&\;\; \\
    $R_{18}$&\;$-2.0\times10^{-3}$\; &\;&\;\; \\
   \hline
&&&      \\[-6pt]
    $C_{1}$&\;$-5.6\times10^{-2}$\; &\;$C_{2}$\;&\;\;\;\;$1.45\times10^{-1}$\; \\
    $C_{2}$&\;$-4.8\times10^{-2}$\; &\;$C_{6}$\;&\;\;\;\;$6.0\times10^{-2}$\; \\
   $C_{3}$&\;\;\;\;$1.2\times10^{-1}$\; &\;$C_{7}$\;&\;$-8.0\times10^{-3}$\; \\
   $C_{4}$&\;\;\;\;$6.0\times10^{-3}$\; &\;$C_{8}$\;&\;\;\;\;$1.62\times10^{-2}$\; \\
    $C_{5}$&\;\;\;\;$8.6\times10^{-3}$\; &\;\;&\;\; \\
    $C_{6}$&\;$-2.0\times10^{-2}$\; &\;\;&\;\; \\
    $C_{7}$&\;$-6.0\times10^{-4}$\; &\;\;&\;\; \\
    $C_{8}$&\;$-2.6\times10^{-3}$\; &\;\;&\;\; \\
    \hline
    \hline
  \end{tabular}
\end{table}  

\subsection{Intervalley scattering}

\subsubsection{Inter--$\Gamma$-$L$ valley scattering}
In this part we investigate the inter--$\Gamma$-$L$ valley
electron-phonon scattering. As
  pointed out by Li {\em et al.},\cite{Li-Ge3} the zeroth-order scattering matrix elements
of both the spin-conserving and spin-flip processes exist. Thus we
first focus on the
  analysis of the zeroth-order contributions with subgroup
 techniques.\cite{Lax-Selection1,Lax2} We then
   investigate the other lowest-order non-vanishing wave-vector
   dependence of the scattering induced by each
   phonon branch with the method of invariants.\cite{Bir1} 

Obviously the scatterings to the four $L$
  valleys are equivalent in unstrained bulk Ge except for the
coordinate rotation. Here we  consider the
$(\pi/a)(1,1,1)$ valley and choose the [111] coordinate system. The
lowest conduction bands at the $\Gamma$ point are basis functions of the $\Gamma^-_7$ irreducible representation,\cite{PeterYu1}
\begin{equation}\label{eigenG}
\Gamma^{-}_{7(1)}=\Gamma^{-}_2\up,\quad\quad\Gamma^{-}_{7(2)}=\Gamma^{-}_2\down,
\end{equation}
where the spin eigenstates are along the $\hat{z}$ direction. The
  eigenstates of conduction band minima at the $L$ point are given in
  Eq.\,(\ref{eigenL}), which are spin-mixed. Besides, the phonon states at the $L$ point,
  which correspond the inter--$\Gamma$-$L$ valley scattering, are
  basis functions of $L_3$, $L_{2'}$, $L_1$
  and $L_{3'}$ irreducible representations.\cite{PeterYu1} 

The symmetry of the inter--$\Gamma$-$L$ valley scattering is described
  by the $D_{3d}$ point group,\cite{Group1,PeterYu1} where the
symmetry of the wave-vectors is given by
\begin{equation}\label{wavevector}
K_z\sim L_{2'},\; (K_x,K_y)\sim L_{3'},\;q_z\sim L_{2'},\; (q_x,q_y)\sim L_{3'}.
\end{equation}
Here ${\bf K}={\bf k}^{\prime}-{\bf k}_L+{\bf k}-{\bf k}_{\Gamma}$
and ${\bf q}={\bf k}^{\prime}-{\bf k}_L-({\bf k}-{\bf k}_{\Gamma})$ with
${\bf k}_{\Gamma}=(\pi/a)(0,0,0)$ and ${\bf
  k}_{L}=(\pi/a)(1,1,1)$.   By subgroup techniques, 
we obtain a general selection rule,\cite{Lax-Selection1,Lax2}
\begin{equation}
  L^+_6\otimes\Gamma^{-}_7=L_{1'}+L_{2'}+L_{3'}.
 \end{equation}
Thus only the zeroth-order contributions from the $L_{2'}$ and
$L_{3'}$ phonons exist.

We then give a more detailed symmetry analysis 
on the zeroth-order contributions. There are various 
electron-phonon mechanisms as the
$L^{+}_6$, $\Gamma^-_7$ and $L_{3'}$ states are all two-fold
degenerate. In each particular case, we perform the operators in
$D_{3d}$ point group on
the initial, final and phonon states. We find that many elements are
forbidden,  and there are relations between the nonvanishing elements
\begin{eqnarray}
    \lan\Gamma^{-}_{7(1)}|V_{L_{2'}}|L^{+}_{6(1)}\ran&\stackrel{ C_{2(1)}}{=}\lan\Gamma^{-}_{7(2)}|V_{L_{2'}}|L^{+}_{6(2)}\ran,\label{G_L_L2}\\
    \lan \Gamma^{-}_{7(1)} |V_{L_{3'(2)}} | L^{+}_{6(2)} \ran
    &\stackrel{ C_{2(1)}}{=}-\lan\Gamma^{-}_{7(2)} |V_{L_{3'(1)}} |L^+_{6(1)} \ran \label{G_L_L3c},  
\end{eqnarray}
in which $C_{2(1)}$ stands for the rotation around the $\hat{x}$ direction
for $\pi$. $L_{3'(1)}$ and $L_{3'(2)}$ [$L^{+}_{6(1)}$ and $L^{+}_{6(2)}$] are basis functions of the $L_{3'}$ ($L^+_6$)
irreducible representation, which are given in Appendix~\ref{appA}.
 One can see that for spin eigenstates along the $\hat{z}$ direction,
  only the $L_{2'}$ ($L_{3'}$) phonon branch
  contributes to the spin-conserving (spin-flip)  scattering,
  which is in agreement with the selection rules in the previous work.\cite{Li-Ge3}  With Eq.\;(\ref{G_L_L3c}) and the time-revesal symmetry [See Eqs.\;(\ref{matrix}) and
    (\ref{time_reversal})], we obtain an additional limit on the $L_{3'}$
    contribution 
\begin{eqnarray}
&&B_{{\bf k}_{\Gamma},{\bf k}_{L};x}(L_{3'})-i B_{{\bf k}_{\Gamma},{\bf k}_{L};y}(L_{3'})\nonumber\\&&=-B_{{\bf k}_{\Gamma},{\bf k}_{L};x}(L_{3'})-i B_{{\bf k}_{\Gamma},{\bf k}_{L};y}(L_{3'}).
\end{eqnarray}
 Thus the inter--$\Gamma$-$L$ valley scattering matrix can be expressed as
\begin{eqnarray}
\hat{M}_{{\bf k}_{\Gamma},{\bf k}_{L}}=  &
  \setlength{\arraycolsep}{2.4mm}
  \left(\begin{array}{cccc}
    A_{{\bf k}_{\Gamma},{\bf k}_{L}}(L_{2'})         & -i B_{{\bf k}_{\Gamma},{\bf k}_{L};y}(L_{3'(2)})        \\
    i B_{{\bf k}_{\Gamma},{\bf k}_{L};y}(L_{3'(1)})  & A_{{\bf k}_{\Gamma},{\bf k}_{L}}(L_{2'})
  \end{array}\right),\nonumber\\
\label{H_G_L}
\end{eqnarray}
with
\begin{equation}
B_{{\bf k}_{\Gamma},{\bf k}_{L};y}(L_{3'(2)})=B_{{\bf k}_{\Gamma},{\bf k}_{L};y}(L_{3'(1)})=B_{{\bf k}_{\Gamma},{\bf k}_{L};y}(L_{3'}),
\end{equation}
where $A_{{\bf k}_{\Gamma},{\bf k}_{L}}(L_{2'})$ [$B_{{\bf k}_{\Gamma},{\bf
      k}_{L};y}(L_{3'})$] is purely real
(imaginary). The coefficients are given  in
Table~\ref{table_G_L}. It should be noted that the two non-diagonal terms come from two
different $L_{3'}$ basis functions separately, which makes the
spin-orientation dependence of scattering matrix in this case different from the general form
in Eq.\;(\ref{orientation}). 

Now we  consider the spin-orientation dependence of the
inter--$\Gamma$-$L$ valley scattering, which is given by 
\begin{eqnarray}
|M_{{\bf k}_{\Gamma},{\bf k}_{L};\up,\up}|^2&=&|M_{{\bf k}_{\Gamma},{\bf
    k}_{L};\up,\up}(L_{2'})|^2+|M_{{\bf k}_{\Gamma},{\bf
    k}_{L};\up,\up}(L_{3'})|^2\nonumber\\
&=&|A_{{\bf
    k}_{\Gamma},{\bf k}_{L}}(L_{2'})|^2\nonumber\\&&\mbox{}\hspace{-0.08cm}+\left|-\fr{i}{2}\sin{\theta}{e^{-i\phi}}B_{{\bf k}_{\Gamma},{\bf
      k}_{L};y}(L_{3'(2)})\right|^2\nonumber\\&&\mbox{}\hspace{-0.08cm}+\left|-\fr{i}{2}\sin{\theta}{e^{i\phi}}B_{{\bf k}_{\Gamma},{\bf
      k}_{L};y}(L_{3'(1)})\right|^2\nonumber\\
&=&|A_{{\bf  k}_{\Gamma},{\bf k}_{L}}(L_{2'})|^2+\fr{1}{2}\sin^2{\theta}|B_{{\bf
      k}_{\Gamma},{\bf k}_{L};y}(L_{3'})|^2,\nonumber\\\label{G_L_spin1}\\
|M_{{\bf k}_{\Gamma},{\bf k}_{L};\up,\down}|^2&=&|M_{{\bf
    k}_{\Gamma},{\bf k}_{L};\up,\down}(L_{2'})|^2+|M_{{\bf
    k}_{\Gamma},{\bf k}_{L};\up,\down}(L_{3'})|^2\nonumber\\
&=& \left|i\sin^2{\fr{\theta}{2}}e^{i\phi}B_{{\bf k}_{\Gamma},{\bf
      k}_{L};y}(L_{3'(2)})\right|^2\nonumber\\&&\mbox{}\hspace{-0.08cm}+\left|i\cos^2{\fr{\theta}{2}}e^{-i\phi}B_{{\bf k}_{\Gamma},{\bf
      k}_{L};y}(L_{3'(1)})\right|^2\nonumber\\
&=&\fr{1+\cos^2{\theta}}{2}|B_{{\bf k}_{\Gamma},{\bf
      k}_{L};y}(L_{3'})|^2,\label{G_L_spin2}
\end{eqnarray}
where $\up$ ($\down$) is the spin eigenstate along an arbitrary
direction
$\hat{n}=(\sin{\theta}\cos{\phi},\sin{\theta}\sin{\phi},\cos{\theta})$.  
One can see that generally both 
the $L_{2'}$ and $L_{3'}$ phonons are involved in the
  spin-conserving scattering and only the $L_{3'}$ one contributes to
  the spin-flip process. The spin-flip scattering shows obvious
spin-orientation dependence, where the scattering is strongest
 (weakest) with the spin eigenstates along (perpendicular to) the
$\hat{z}$ direction. This anisotropy will be weakened when considering the other
three channels of inter--$\Gamma$-$L$ valley scatterings, which can be
  obtained by the coordinate rotations. Numerically, 
 the spin-conserving coefficient $|A_{{\bf
    k}_{\Gamma},{\bf k}_{L}}(L_{2'})|$ is much larger (about 50 times)
than the spin-flip one $|B_{{\bf
    k}_{\Gamma},{\bf k}_{L};y}(L_{3'})|$ (See Table~\ref{table_G_L}),
and therefore the
spin-conserving scattering is nearly isotropic.
\begin{table}[htb]
  \caption{The coefficients in the inter--$\Gamma$-$L$ valley
      electron-phonon scattering matrix and the first- and second-order
 contributions to the spin-flip process
in  the  spherical band-approximation.
$A_{{\bf  k}_{\Gamma},{\bf k}_{L}}$ and $B_{{\bf k}_{\Gamma},{\bf
    k}_{L};y}$ represent the values for the zeroth-order contributions 
to the scattering
   matrix.  $R_i$ and $\Xi_i$ are the coefficients of the first-order
  terms in the scattering matrices, and $C_i$ denote the parameters
    of the second-order ones.  $D_1$ and $D_2$
  are the parameters of the first- and second-order contributions 
to the spin-flip
 process in the spherical-band approximation,
with the spin eigenstates along the [111] direction.}
  \label{table_G_L}
 \begin{tabular}{ll|ll}
    \hline
    \hline
    &eV/nm\;&  &\;\;eV \\ [4pt] 
    \hline 
    &&&      \\[-6pt]
    $A_{{\bf k}_{\Gamma},{\bf k}_{L}}(L_{2'})$&$18.21$&\;$R_1$&$0.73$\;  \\[4pt]
    $iB_{{\bf k}_{\Gamma},{\bf k}_{L};y}(L_{3'})$&\;\,$0.35$&\;$R_2$&\hspace{-0.25cm}$-0.34$\;  \\[4pt]
    && \;$R_3$&\hspace{-0.25cm}$-2.15\times10^{-2}$\;\\[4pt]
   &&\;$R_4$&$2.46\times10^{-2}$\; \\[4pt]
   &&\;$R_{5}$&$37.64\times10^{-2}$\; \\[4pt]
  &&\;$R_{6}$&\hspace{-0.25cm}$-38.23\times10^{-2}$\;  \\[4pt]
  &&\;$R_{7}$&$1.66$\;  \\[4pt]
  &&\;$R_{8}$&\hspace{-0.25cm}$-2.06$\;  \\[4pt]
     \hline
    \hline
&\;\;\;\;\;\;eV&&\;\;\; eV$\cdot$\AA  \\ [4pt] 
 \hline 
    && &     \\[-6pt]
$\Xi_1$&\;\;\;\;$12.0$ &\;$C_1$&\;\;\;\;12.7\\[4pt]
$\Xi_2$&\;$-9.15$\; &\;$C_2$&\;\;\;\;4.54\\[4pt]
 $\Xi_3$&\;$-30.5$\; &\;$C_3$&\;\;\;\;6.04\\[4pt]
$\Xi_4$&\;\;\;\;$20.5$\;&\;$C_4$&\;\;\;\;$2.51$\\[4pt]
$D_{1}(L_3)$&\;\;\;\;$2.79\times10^{-3}$\;&\;$C_5$& \;$-15.5$ \\[4pt]
$D_{1}(L_1)$&\;\;\;\;$2.14\times10^{-1}$\;&\;$C_6$& \;\;\;\;4.16
 \\ [4pt]
&&\;$C_7$& \;\;\;\;7.08 \\[4pt]
&&\;$C_8$& \;$-3.98$ \\[4pt]
&&\;$D_{2}(L_{2'})$&\;\;\;\;0.54 \\[4pt]
\hline\hline
  \end{tabular}
\end{table} 

As the zeroth-order terms of the contributions of the $L_3$/$L_1$ phonon
and the spin-flip process of the $L_{2'}$ phonon are absent, we further
derive the lowest-order non-vanishing wave-vector
dependence of these contributions with the
 method of invariants. From the space-inversion symmetry of the
electron/phonon states and the wave-vectors [See Eq.\;(\ref{wavevector})], one 
finds that the $L_3$/$L_1$ ($L_{2'}$) phonons
only contribute to the odd- (even-)order terms in the scattering
matrix. For the $L_3$/$L_1$
   phonons, the first-order terms exist in both the spin-conserving
   and spin-flip processes, given by
\begin{subequations}\label{L3_D1}
\begin{align}
A(L_{3x})&=-i(\Xi_1q_y+\Xi_2K_y),\label{L310} \\ B_z(L_{3x})&=R_1q_x+R_2K_x,\label{L31z}\\
B_{x}(L_{3x})&=-(R_3q_y+R_4K_y-R_5q_z-R_6K_z),\label{L31x}\\
B_{y}(L_{3x})&=-R_3q_x-R_4K_x,\label{L31y}\\
A(L_{3y})&=i(\Xi_1q_x+\Xi_2K_x), \label{L320}\\
B_z(L_{3y})&=R_1q_y+R_2K_y,\label{L32z}\\
B_{x}(L_{3y})&=-R_3q_x-R_4K_x,\label{L32x}\\
B_{y}(L_{3y})&=R_3q_y+R_4K_y+R_5q_z+R_6K_z,\label{L32y}
\end{align}
\end{subequations}
and 
\begin{subequations}\label{L1_D1}
\begin{align}
A(L_1)&=-i(\Xi_3q_z+\Xi_4K_z),\label{L1_0}\\
B_x(L_1)&=R_7q_y+R_8K_y,\label{L1_x}\\
B_y(L_1)&=-(R_7q_x+R_8K_x),\label{L1_y}
\end{align}
\end{subequations}
with $L_{3x}\sim -zy$ and $L_{3y}\sim zx$ being the basis functions of
$L_3$ irreducible representation. For the $L_{2'}$ phonon, the
second-order terms are nonvanishing in both the spin-conserving
and spin-flip processes. Nevertheless, its contribution to the
spin-conserving process is negligible 
compared to the zeroth-order contribution 
$A_{{\bf  k}_{\Gamma},{\bf k}_{L}}(L_{2'})$. As there is no zeroth-order
term in the spin-flip process, the 
second-order spin-flip terms become important and read
\begin{equation}\label{L2_z}
B_z(L_{2'})=iC_1({\bf K}\times{\bf q})_z,
\end{equation}
and 
\begin{equation}\label{L2_xy}
[B_x(L_{2'}),B_y(L_{2'})]=i\sum\limits_{j=2}^8C_j(S^{(j)}_x,S^{(j)}_y),
\end{equation}
with
\begin{subequations}\label{L2_xy_2}
\begin{align}
(S^{(2)}_x,S^{(2)}_y)&=K_z(K_y,-K_x),\\
(S^{(3)}_x,S^{(3)}_y)&=K_z(q_y,-q_x),\\
(S^{(4)}_x,S^{(4)}_y)&=q_z(q_y,-q_x),\\
(S^{(5)}_x,S^{(5)}_y)&=q_z(K_y,-K_x),\\
(S^{(6)}_x,S^{(6)}_y)&=(K^2_x-K^2_y,-2K_xK_y),\\
(S^{(7)}_x,S^{(7)}_y)&=(q^2_x-q^2_y,-2q_xq_y),\\
(S^{(8)}_x,S^{(8)}_y)&=(K_xq_x-K_yq_y,-K_xq_y-K_yq_x).
\end{align}
\end{subequations}
The coefficients in  Eqs.\;(\ref{L3_D1})-(\ref{L2_xy_2}) are given in
 Table~\ref{table_G_L} and  the spin-orientation dependence of the
 first- and second-order scattering matrices
 are expressed by Eq.\;(\ref{orientation}). Note we have again omitted the
   subscripts ${\bf k}$ and ${\bf k}'$ in
 Eqs.\;(\ref{L3_D1})-(\ref{L2_xy_2}). One can see that the first-order
 spin-conserving  terms ($\Xi_i$) are much larger than the spin-flip
 ones ($R_i$).  We also show the  first-order
 contributions (of the $L_3$ and $L_1$ phonons) and second-order
 one (of the $L_{2'}$ phonon) to the spin-flip processes of the
 inter--$\Gamma$-$L$ scattering in
the spherical-band approximation in Table~\ref{table_G_L} for completeness.

\subsubsection{Inter--$L$-$L$ valley scattering}
Recently, Tang {\em et al.}\cite{Tang-Ge1} and Li {\em et
  al.}\cite{Li-Ge2} demonstrated the inter-$L$ valley electron-phonon
scattering with the $X$-point phonons (belong to $X_3$, $X_1$
and $X_4$ irreducible representations) involved.\cite{Group1,PeterYu1}
Tang {\em et  al.}\cite{Tang-Ge1} gave the general selection rules for the
zeroth-order scattering matrix element and calculated the average
values of both the zeroth-
and first-order contributions, where the zeroth-order $X_3$-phonon
  contribution is forbidden. Li {\em et
  al.}\cite{Li-Ge2} further derived the spin-orientation
dependence and complete scattering matrices for the zeroth-order
contributions, which were expressed with coefficients
$D_{X_1,m}$, $D_{X_1,s}$ and $D_{X_4,s}$. Here $D_{X_1,m}$ corresponds
to the spin-conserving process only and the $D_{X_1,s}$ and
$D_{X_4,s}$ are coefficients for both the spin-flip and
spin-conserving processes. In this work we additionally derive the
  explicit scattering matrix for the first-order terms of the $X_3$-phonon
  contribution with the method of invariants, which are shown to be
  non-negligible for the intrinsic electron spin relaxation in Ge in low
  temperature due to the relative low $X_3$-phonon energy.\cite{Li-Ge2}

As the scatterings between the four $L$ valleys are equivalent
  except for the coordinate rotations, we consider the
  $(\pi/a)(1,1,1)\leftrightarrow(\pi/a)(1,1,-1)$ case only and take
  the crystallographic frame for simplicity. The space symmetry of this
  scattering is described by one subgroup of the $G^{2}_{32}$ group (See
  Table~\ref{characterX}),\cite{PeterYu1,Lax-Selection1,Tang-Ge1,Li-Ge2}
  where the 2-fold $X_3$ irreducible representation in $G^{2}_{32}$ group can be deduced into
  two one-dimensional (1D) representations and the wave-vectors
  and Pauli matrices belong to 1D representations also. Based on the
  space symmetry of the initial/final states, phonon states,
  wave-vectors and Pauli matrices (shown in Table~\ref{characterX}) and
  the time-reversal symmetry, we construct the complete scattering matrix,
  \begin{subequations}
    \begin{align}
      A(X^-_{3})&=-i\Xi_1(q_x-q_y),\\
      B_z(X^-_{3})&=R_1(K_x+K_y)+R_2q_z,\\
      B_x(X^-_{3})&=R_3(q_x+q_y)+R_4(q_x-q_y)+R_5K_z,\\
      B_y(X^-_{3})&=R_3(q_x+q_y)-R_4(q_x-q_y)+R_5K_z.
    \end{align}
  \end{subequations}
   Here $X^{\pm}_{3}\sim(\sin{\frac{4\pi x}{a}}\pm\sin{\frac{4\pi
      y}{a}})(\cos{\frac{2\pi z}{a}}\pm\sin{\frac{2\pi z}{a}})$ are the
  basis functions of the $X_3$ irreducible representation and the
  $X^{+}_{3}$-phonon contribution is forbidden by the space symmetry (See
  Table~\ref{characterX}). ${\bf K}={\bf k}^{\prime}-{\bf k}_{\rm
    i}+{\bf k}-{\bf k}_{\rm f}$
and ${\bf q}={\bf k}^{\prime}-{\bf k}_{\rm i}-({\bf k}-{\bf k}_{\rm f})$ with
${\bf k}_{\rm i}=(\pi/a)(1,1,1)$ and ${\bf
  k}_{\rm f}=(\pi/a)(1,1,-1)$. The coefficients are given in
  Table~\ref{table_T_L} and the spin-orientation dependence is given by Eq.\;(\ref{orientation}). One can see that the $X_3$ phonons
  contribute to both the spin-conserving and spin-flip
  processes of the inter--$L$-$L$ scattering, and the spin-conserving
  one is much larger (about 2 orders of magnitude) than the spin-flip
  one. We also calculate the spin-flip first-order contribution of the 
  $X_3$ phonons in the
  spherical-band-approximation and  the zeroth-order contributions of the 
  $X_1$ and $X_4$  phonons for the sake of completeness (listed in
  Table~\ref{table_T_L}),  where
we take the same notations for the zeroth-order scattering
matrices as those in Li's work.\cite{Li-Ge2} The
parameters are in the same order with those in the previous
works.\cite{Tang-Ge1,Li-Ge2}

     \begin{table}[htb]
        \caption{The space symmetry of the final/initial electron states,
          phonon states, wave-vectors and Pauli matrices  in one
          subgroup of the $G^2_{32}$ group for
          the $(\pi/a)(1,1,1)\leftrightarrow(\pi/a)(1,1,-1)$ scattering.\cite{Lax-Selection1,Li-Ge2,PeterYu1} Here
          $\tilde{\sigma}_{\pm}=\frac{1}{\sqrt{2}}(\sigma_{x}\pm\sigma_{y})$,
          $\tilde{k}_{\pm}=\frac{1}{\sqrt{2}}(k_{x}\pm k_{y})$ and
          $\tau=(a/4)(1,1,1)$. $X^{\pm}_{3}$ are the basis functions of
          the $X_{3}$ irreducible representations in the $G^{2}_{32}$
          group, which are given in main text.}
        \label{characterX}
        \begin{tabular}{l|lllllllllll}
          \hline
          \hline
          &&&&&&&&&&&\\[-6pt]
          \;\;\;\;\;g &\;\;\;$I$\;&\;$\tilde{\sigma}_{\pm}$\;&\;$\sigma_z$\;&\;$\tilde{k}_{\pm}$\;&\;$k_{z}$\;&\;$X^{\pm}_{3}$&\;\;$L_{1}$\;&\;$L_{1t}$
          \\[4pt]
          \hline
          &&&&&&&&&&&\\[-8pt]
          $(E|0)$\;      &\;\;\;1&\;\;\;1&\;\;\;1 &\;\;\;1   &\;\;\;1 &\;\;\;1   &\;\;\;1 &\;\;\;1   \\[4pt]
          \hline
          &&&&&&&&&&&\\[-8pt]
          $(C_{2xy}|\tau)$\;  &\;\;\;1 &$\pm1$     &$-1$&$\pm1$  &$-1$    &\;\;\;1   &$-L_{1t}$
          &\;\;\;$L_{1}$   \\[4pt]
          \hline
          &&&&&&&&&&&\\[-8pt]
          $(C_{2x\bar{y}}|\tau)$\;  &\;\;\;1 &$\mp1$  &$-1$ &$\mp1$           &$-1$  &$-1$ &\;\;\;1
          &$-1$        \\[4pt]
          \hline
          &&&&&&&&&&&\\[-8pt]
          $(C_{2z}|0)$\;  &\;\;\;1 &$-1$     &\;\;\;1 &$-1$  
          &\;\;\;1 &$-1$  &\;\;\;$L_{1t}$ &\;\;\;$L_{1}$   \\[4pt]
          \hline
          &&&&&&&&&&&\\[-8pt]
          $(i|\tau)$\;  &\;\;\;1 &\;\;\;1  &\;\;\;1 &$-1$   
          &$-1$ &$\mp1$   &\;\;\;1 &$-1$  \\[4pt]
          \hline
          &&&&&&&&&&&\\[-8pt]
          $(\rho_{xy}|0)$ \; &\;\;\;1 &$\pm1$   &$-1$ &$\mp1$   
          &\;\;\;1 &$\mp1$ &\;\;\;$L_{1t}$ &\;\;\;$L_{1}$  \\[4pt]
          \hline
          &&&&&&&&&&&\\[-8pt]
          $(\rho_{x\bar{y}}|0)$\;  &\;\;\;1 &$\mp1$   &$-1$ &$\pm1$ 
          &\;\;\;1 &$\pm1$  &\;\;\;1 &\;\;\;1    \\[4pt]
          \hline
          &&&&&&&&&&&\\[-8pt]
          $(\rho_{z}|\tau)$\;  &\;\;\;1 &$-1$   &\;\;\;1 &\;\;\;1 
          &$-1$ &$\pm1$  &$-L_{1t}$ &\;\;\;$L_{1}$  \\[4pt]
          \hline
          \hline
        \end{tabular}
      \end{table}

\begin{table}[htb]
  \caption{The coefficients in the inter--$L$-$L$ valley
      electron-phonon scattering matrix.  In the first two
      columns, the $\Xi_i$ and $R_i$s  stand for the parameters of the first-order
      scattering matrices of $X_3$ phonons and $D_{1}(X_3)$
      is that for the first-order $X_3$-phonon spin-flip scattering element in the
      spherical-band-approximation with the spin eigenstates along the
  [001] direction. The last two columns show the
      coefficients of the zeroth-order contributions to the scattering
      matrix.\cite{Li-Ge2}}
  \label{table_T_L}
   \begin{tabular}{ll|ll}
    \hline
    \hline
    &&&      \\[-6pt]
    \;&\quad\;  eV\quad &&\;\; eV/nm\; \\ [4pt]  
    \hline
    \hline
    &&&      \\[-6pt]
     \;  $\Xi_{1}$&\;\;\;\;$6.49$&\;$D_{X_1,s}$&\quad$0.18$     \\       [4pt]
     \;  $R_{1}$&\;$-3.65\times10^{-2}$\;&\;$D_{X_4,s}$&\quad$0.66$\\[4pt]
    \;  $R_{2}$&\;\;\;\;$1.43\times10^{-1}$\;&\;$D_{X_1,m}$&\quad$6.56$  \\ [4pt]
  \;  $R_{3}$&\;$-4.58\times10^{-2}$\;& &\;\\ [4pt]
  \;   $R_{4}$&\;\;\;\;$5.17\times10^{-2}$\;& &\;\\ [4pt]
  \;   $R_{5}$&\;\;\;\;$8.68\times10^{-2}$\;& &\;\\ [4pt]
  \;    $D_{1}(X_3)$&\;\;\;\;$6.71\times10^{-2}$\;  &&\;     \\       [4pt]
    \hline
    \hline
  \end{tabular}
\end{table}

\subsection{Elliott and Yafet mechanisms}
In this part we analyze the  Elliott\cite{Elliott1} and
Yafet\cite{Yafet1} mechanisms  in the spin-flip part of the
electron-phonon scattering, where the electron-phonon interaction can be generally expressed as\cite{Yafet1,Dery_Si1}
\begin{equation}
H_{\rm ep}=\sum_{{\bf q},\lambda}{\bf u}_{{\bf
    q},\lambda}\cdot\nabla[V_0({\bf r})\hat{I}+\fr{\hbar}{4m^2_0c^2}(\nabla
  V_0({\bf r})\times{\bf p})\cdot{\bm \sigma}].\label{Hep_EY}
\end{equation}
Here  ${\bf u}_{{\bf q},\lambda}$ is the phonon displacement vector
and ${\bf q}$ (${\lambda}$) denotes the phonon wave vector (phonon
mode). $V_0({\bf r})\hat{I}$ represents the bare potential and $\fr{\hbar}{4m^2_0c^2}(\nabla
  V_0({\bf r})\times{\bf p})\cdot{\bm \sigma}$ stands for the spin-orbit
  coupling. It should be noted that the bare potential and the spin-orbit coupling, which correspond to the Elliott and Yafet processes separately,
share the same symmetry in all the point groups of the crystal. 
Therefore their scattering matrices have the same analytical 
forms with the total electron-phonon scattering matrices,
 which are the sum of the Elliott
  and Yafet ones. 

For the spin-flip intravalley scattering, the wave-vector order analysis is
  clear in the centrosymmetric crystal.\cite{Elliott1,Yafet1,Li-Ge2} Our analytical results
   are [See Eqs.\;(\ref{G_Sz}), (\ref{L^{+}_Bz}) and
    (\ref{L^{+}_Bxy})]  in agreement with the previous statements.\cite{Elliott1,Yafet1,Li-Ge2,Dery_Si1} In the
  AC-phonon induced scattering,  the first-order ($K_l$) contributions to the Elliott
  and Yafet processes cancel each other completely and the
  third-order terms ($K_lq_mq_n$)  remain. While in the OP case, the
  leading-order terms of both the Elliott and Yafet mechanisms are
  second-order ($K_lq_m$)  and
  there is no such perfect cancellation. Here $l,m,n\in\{x,y,z\}$.  In
  our numerical calculation, the first-order contributions to the
  Elliott and Yafet processes in the AC spin-flip scattering are much
  larger (more than three orders of magnitude) than the total third-order spin-flip scattering matrix
  elements; and both the Elliott and Yafet terms in OP cases are in the same order with the total spin-flip matrix elements.

\begin{table}[htb]
  \caption{The coefficients for the zeroth-order contributions of the Elliott
      and Yafet mechanisms in the inter--$\Gamma$-$L$ and inter-$L$ valley
      electron-phonon scattering matrices. The superscript ``E''
      (``Y'') stands for the Elliott (Yafet)
mechanism. The first two lines represent the parameters of the
spin-flip inter--$\Gamma$-$L$ scattering and the following four lines
stand for those in the inter-$L$ case. One can see that the
coefficients in Tables~\ref{table_G_L} and \ref{table_T_L} are
sums of the corresponding Elliott and Yafet ones in this table.}
  \label{table_EY}
  \begin{tabular}{ll}
    \hline
    \hline
    &     \\[-6pt]
    &\;  eV/nm\; \\ [4pt]  
    \hline
    \hline
    &      \\[-6pt]
    $iB^{\rm E}_{{\bf k}_{\Gamma},{\bf k}_{L};y}(L_{3'})$&\quad$-0.10$ \\      [4pt]
    $iB^{\rm Y}_{{\bf k}_{\Gamma},{\bf k}_{L};y}(L_{3'})$&\quad\;\;\;$0.45$  \\      [4pt]
    $D^{\rm E}_{1,s}(X_1)$&\quad$-0.08$\\        [4pt]
    $D^{\rm Y}_{1,s}(X_1)$&\quad\;\;\;$0.26$ \\        [4pt]
    $D^{\rm E}_{4,s}(X_4)$&\quad\;\;\;$0.25$\\        [4pt]
    $D^{\rm Y}_{4,s}(X_4)$&\quad\;\;\;$0.41$\\        [4pt]
     \hline
    \hline
  \end{tabular}
\end{table}

We then turn to the Elliott-Yafet
mechanism  in the zeroth-order spin-flip inter--$\Gamma$-$L$ and inter-$L$ valley
scatterings, where our pseudo-potential calculation of the Elliott-Yafet coefficients are listed in
Table~\ref{table_EY}. Generally, the Elliott and Yafet  contributions
 are in the same order with those of the total scattering
matrices, which are simply sums of the corresponding Elliott-Yafet
    coefficients (See Tables~\ref{table_G_L} and \ref{table_T_L}).
 We further analyze the zeroth-order contribution of these 
two mechanisms  with the $\Gamma$-
and $L$-point ${\bm k}\cdot{\bm p}$ eigenstates [See Eqs.\,(\ref{eigenG}) and
  (\ref{eigenL})], and give the explicit expressions with the
single-group basis functions. For the inter--$\Gamma$-$L$ valley scattering, the
Elliott-Yafet matrix elements can be expressed as 
\begin{eqnarray}
iB^{\rm E}_{{\bf k}_{\Gamma},{\bf k}_{L};y}(L_{3'})&=&-\fr{c}{A}\lan
  \Gamma^{-}_{2}\up|V_{L_{3'(2)}}^{\rm E}| L_{3(1)}\up\ran,\\
iB^{\rm Y}_{{\bf k}_{\Gamma},{\bf k}_{L};y}(L_{3'})&=&\fr{1}{A}[ \lan\Gamma^{-}_{2}\up|V_{L_{3'(2)}}^{\rm
     Y}|L_1\down\ran\nonumber\\&&\mbox{}-c\lan
  \Gamma^{-}_{2}\up|V_{L_{3'(2)}}^{\rm
     Y}|L_{3(1)}\up\ran],
\end{eqnarray}
where $\up$ ($\down$) is the spin eigenstate along the [111] direction
and the superscript ``E'' (``Y'') denotes the Elliott (Yafet) mechanism.  One can see that the coupling between the lowest and upper conduction
bands ($L_1$ and $L_3$) at the $L$-point is critical to the  Elliott contribution, which is
non-negligible in the total scattering matrix element. Here
the Yafet contribution is about 4 times larger then the Elliott one
(See Table~\ref{table_EY}).  For the
inter-$L$ valley scattering,  we consider the scattering between
${\bf k}_{L_t}=\fr{\pi}{a}(1,1,-1)$ and ${\bf
  k}_{L}=\fr{\pi}{a}(1,1,1)$ and take the spin eigenstates along the [001]
direction. The Elliott-Yafet matrix elements are
given by
\begin{eqnarray}
D^{\rm
  E}_{1,s}(X_1)&=&-\fr{\sqrt{2}c}{2A^2}[\lan L^{t}_1\up^{\prime}|V^{\rm
E}_{X_1}|(L_{3y}+i\cos{\theta}L_{3x})\up^{\prime}\ran\nonumber\\
&&\mbox{}-\lan (L^{t}_{3y}+i\cos{\theta}L^{t}_{3x}) \down^{\prime}|V^{\rm
E}_{X_1}|L_1\down^{\prime}\ran],\\\label{DEX4}
D^{\rm
  Y}_{1,s}(X_1)&=&-\fr{\sqrt{2}c}{2A^2}[\lan L^{t}_1\up^{\prime}|V^{\rm
Y}_{X_1}|(L_{3y}+i\cos{\theta}L_{3x})\up^{\prime}\ran\nonumber\\
&&\mbox{}-\lan (L^{t}_{3y}+i\cos{\theta}L^{t}_{3x}) \down^{\prime}|V^{\rm
Y}_{X_1}|L_1\down^{\prime}\ran]\nonumber\\
&&\mbox{}-\fr{\sqrt{2}e^{i\phi}}{2A^2}[\lan L^{t}_1\up^{\prime}|V^{\rm
Y}_{X_1}|L_1\down^{\prime}\ran\nonumber\\
&&\mbox{}+ic\sin{\theta}(\lan L^{t}_{3x} \up^{\prime}|V^{\rm
Y}_{X_1}|L_1\down^{\prime}\ran\nonumber\\&&\mbox{}+\lan L^{t}_1\up^{\prime} |V^{\rm
Y}_{X_1}| L_{3x}\down^{\prime}\ran)],\label{DYX4}
\end{eqnarray}
and
\begin{eqnarray}
D^{\rm E}_{1,s}(X_4)&=&-\fr{i}{A^2}[ic\sin{\theta}(\lan L^{t}_{3x} \up^{\prime}|V^{\rm
E}_{X_4}|L_1\up^{\prime}\ran\nonumber\\
&&\mbox{}-\lan L^{t}_1\up^{\prime} |V^{\rm
E}_{X_4}| L_{3x}\up^{\prime}\ran)],\\\label{DEX1}
D^{\rm
  Y}_{1,s}(X_4)&=&-\fr{i}{A^2}\{[ic\sin{\theta}(\lan L^{t}_{3x} \up^{\prime}|V^{\rm
Y}_{X_4}|L_1\up^{\prime}\ran\nonumber\\
&&\mbox{}-\lan L^{t}_1\up^{\prime} |V^{\rm
Y}_{X_4}| L_{3x}\up^{\prime}\ran)]\nonumber\\
&&\mbox{} -c[e^{i\phi}\lan L^{t}_1\up^{\prime}|V^{\rm
Y}_{X_4}|(L_{3y}-i\cos{\theta}L_{3x})\down^{\prime}\ran\nonumber\\
&&\mbox{}+e^{-i\phi}\lan(L^{t}_{3y}+i\cos{\theta}L^{t}_{3x})\down^{\prime}|V^{\rm
Y}_{X_4}|L_1\up^{\prime}\ran]\},\nonumber\\\label{DYX1}
\end{eqnarray}
where the superscript ``t'' stands for the wave function at the ${\bf
  k}_{L_t}$ point and $\up^{\prime}$ ($\down^{\prime}$) is the spin eigenstate
along the [001] direction. $\theta=\arccos{\fr{\sqrt{3}}{3}}$ and
$\phi=\fr{\pi}{4}$. We omit the $c^2$ terms in
Eqs.\;(\ref{DEX1})-(\ref{DYX4}) as $c\sim10^{-2}$.  Here the mixing
between the lowest and higher conduction bands leads to the Elliott process
in $D_{1,s}(X_1)$ and both the Elliot and Yafet ones in
  $D_{1,s}(X_4)$. The Elliott and Yafet
  contributions are comparable also in this case (See Table~~\ref{table_EY}).  The scatterings between other $L$ points are
similar to this case and we ignore the discussions in this work.

\section{SUMMARY}

In summary, we have investigated the electron-phonon
  scatterings of the $L$ and $\Gamma$ valleys, and studied the energy spectra near
  the bottom of the conduction bands with the $L$-point ${\bm k}\cdot{\bm p}$
 Hamiltonian in Ge. We first construct a $16\times16$  ${\bm k}\cdot{\bm p}$
 Hamiltonian in the vicinity of the $L$ point with the double-group
basis functions, of which the energy spectra of the lowest three conduction and
highest two valence bands agree with the tight-binding
model ones.  The eigenstates of this Hamiltonian are useful for the 
analysis of the spin-related properties in Ge.

We then study the phonon-induced  electron scatterings of the $L$
  and $\Gamma$ valleys in Ge,  {i.e.}, the intra-$\Gamma$/$L$ valley, inter--$\Gamma$-$L$
valley and inter-$L$ valley scatterings.  Via the symmetry
consideration, we derive the selection rules and compact scattering
matrices in all these cases, among which the
  scattering matrices for the intra-$\Gamma$ valley, inter--$\Gamma$-$L$
valley, OP contribution and the separated TA and LA
contributions of the
intra-$L$ valley scatterings are absent in the literature. We
  show the
lowest-order wave-vector dependence of the
scattering matrices for all the related phonon modes, where the zeroth-order
spin-flip scattering matrix elements are absent (present) in the
intra- (inter-) valley cases. For completeness we also give the
lowest non-zeroth order contributions in the
intervalley cases in the spherical band approximation. The
spin-orientation dependence of the electron-phonon scattering can be
easily obtained with the corresponding scattering matrix.  Our
pseudo-potential calculation provides the 
coefficients of these scattering matrices, and confirms the selection rules and
wave-vector dependence. Finally, we analyze the Elliott and Yafet
mechanisms in these electron-phonon scatterings with the 
${\bm k}\cdot{\bm p}$ eigenstates at the $L$ and $\Gamma$ valleys. This
investigation provides the necessary electron-phonon scatterings for
the study of the  optical orientation of spin
and the hot-electron relaxation in Ge.

\begin{acknowledgments}
This work was supported by the National Basic Research
Program of China under Grant No. 2012CB922002 and the Strategic Priority
Research Program of the Chinese Academy of Sciences under Grant No. XDB01000000.
\end{acknowledgments}

\begin{appendix}
  \section{THE ${\bm k}$$\cdot$${\bm p}$ BASIS FUNCTIONS}\label{appA}
\begin{table}[htb]
 \caption{The basis functions representing 16 $L$-point Bloch states, 4 in the valence band and 12 in the conduction band.}
  \label{tableX}
 \begin{tabular}{l|l|l}
    \hline
    Band state &  Basic functions & $H_{0,jj}$ \\ \hline 
    $|v1\rangle=L^{-}_{4}(L_{3'}^v)$ & $\fr{1}{2}[(x-\mathrm{i}y)\down-(x+\mathrm{i}y)\up]$& $E_{45v}^-$\\
    $|v2\rangle=L^{-}_{5}(L_{3'}^v)$ & $\fr{1}{2}[(x-\mathrm{i}y)\down+(x+\mathrm{i}y)\up]$& $E_{45v}^-$\\
    $|v3\rangle=L^{-}_{6(1)}(L_{3'}^v)$ & $\frac{1}{\sqrt{2}}(x+iy)\down=L_{3'(2)}\down$& $E_{6v}^-$\\
    $|v4\rangle=L^{-}_{6(2)}(L_{3'}^v)$ & $- \frac{1}{\sqrt{2}}(x-iy)\up=-L_{3'(1)}\up$& $E_{6v}^-$\\
    \hline \hline
    $ |c1\rangle=L^{+}_{6(1)}(L_1)$ & $s\up$ & $E_{6c}^+$\\
    $ |c2\rangle=L^{+}_{6(2)}(L_1)$ & $s\down$ & $E_{6c}^+$ \\
    $|c3\rangle=L^{+}_{6(2)}(L_3)$ & $z(x-iy)\up$; $i(x + i y)^2 \up$ & $E_{6c'}^+$\\
    $|c4\rangle=L^{+}_{6(1)}(L_3)$ & $-z(x+iy)\down$; $i(x - i y)^2 \down$ & $E_{6c'}^+$\\
     $|c5\rangle=L^{+}_{4}(L_3)$ & $\frac{1}{\sqrt{2}} z[(x-\mathrm{i}y)\down+(x+\mathrm{i}y)\up]$;& $E_{45c}^+$\\
     & $\frac{i}{\sqrt{2}} [-(x - i y)^2 \up + (x + i y)^2 \down]$&\\
    $|c6\rangle=L^{+}_{5}(L_3)$ & $\frac{1}{\sqrt{2}} z[(x-\mathrm{i}y)\down-(x+\mathrm{i}y)\up]$;& $E_{45c}^+$\\
    & $\frac{i}{\sqrt{2}} [ (x - i y)^2 \up + (x + i y)^2 \down] $& \\
    $|c7\rangle=L^{-}_{4}(L_{3'}^c)$ & {\rm the same as}~~$L^{-}_{4}(L_{3'}^v)$ & $E_{45c}^-$\\
    $|c8\rangle=L^{-}_{5}(L_{3'}^c)$ &  {\rm see}~~ $L^{-}_{5}(L_{3'}^v)$& $E_{45c}^-$ \\
     $|c9\rangle=L^{-}_{6(1)}(L_{3'}^c)$ & {\rm see}~~ $L^{-}_{6(1)}(L_{3'}^v)$ & $E_{6c}^-$\\
    $|c10\rangle=L^{-}_{6(2)}(L_{3'}^c)$ & {\rm see}~~ $L^{-}_{6(2)}(L_{3'}^v)$ & $E_{6c}^-$ \\
    $ |c11\rangle=L^-_{6(1)}(L_{2'})$ & $z\up$ & $E_{6c'}^-$\\
    $|c12\rangle=L^-_{6(2)}(L_{2'})$ & $z\down$ & $E_{6c'}^-$\\
    \hline
  \end{tabular}
\end{table}

In Table~\ref{tableX}, in order to define the transformation matrices for the 16 $L$-point Bloch functions used in our ${\bm
  k}$$\cdot$${\bm p}$ Hamiltonian, we explicitly give in the second column simple examples of the corresponding basis functions. 
The first column  presents the state notation $|v,j\rangle$ with $j=1\dots4$ or $|c,j
\rangle$ ($j = 1\dots 12$), the double-group representation and, in brackets, the corresponding single-group representation. For example, the symbol $L^{-}_{6(2)}(L_{3'}^v)$ means the second state of the representation $L^{-}_6$ originating from the valence-band representation $L_{3'}$. The arrow  $\up$ ($\down$) symbolizes the spin-up (down) eigenstate along the $\hat{z}$
direction. For each of the representations $L_4^+, L_5^+$ we give two
different examples of the basis functions. The notation of the
diagonal energies $H_{0,jj}$ is shown in the third column.
Finally, in construction of the basis functions, we set one of
  the planes $\sigma_v$ contain the axes $y$ and $z$ in the $D_{3d}$ group.

\end{appendix}

\end{document}